\documentclass[pdflatex,sn-mathphys-num]{sn-jnl}
\usepackage[T1]{fontenc}   
\usepackage[utf8]{inputenc}
\pdfoutput=1               

\usepackage{graphicx}%
\usepackage{multirow}%
\usepackage{amsmath,amssymb,amsfonts}%
\usepackage{amsthm}%
\usepackage{mathrsfs}%
\usepackage[title]{appendix}%
\usepackage{xcolor}%
\usepackage{textcomp}%
\usepackage{manyfoot}%
\usepackage{booktabs}%
\usepackage{algorithm}%
\usepackage{algorithmicx}%
\usepackage{algpseudocode}%
\usepackage{listings}%
\usepackage{subcaption}
\usepackage{caption}
\usepackage{graphicx}%
\usepackage{booktabs}
\usepackage{float}
\usepackage{multirow}%
\usepackage{amsmath,amssymb,amsfonts}%
\usepackage{amsthm}%
\usepackage{mathrsfs}%
\usepackage[title]{appendix}%
\usepackage{xcolor}%
\usepackage{textcomp}%
\usepackage{manyfoot}%
\usepackage{booktabs}%
\usepackage{algorithm}%
\usepackage{algorithmicx}%
\usepackage{algpseudocode}%
\usepackage{placeins}
\usepackage{subcaption}
\usepackage{caption}
\usepackage{graphicx}
\usepackage{caption}
\usepackage{amsmath, amssymb}
\usepackage{hyperref}
\usepackage{setspace}
\usepackage{geometry}
\usepackage{float}
\usepackage{booktabs}
\usepackage[most]{tcolorbox}
\usepackage{lmodern} 
\usepackage{etoolbox} 
\usepackage{color}
\usepackage{float}
\usepackage{textcomp}

\definecolor{commentgray}{gray}{0.5}
\definecolor{keywordblue}{rgb}{0.13,0.13,1}
\definecolor{stringred}{rgb}{0.7,0,0}

\lstdefinelanguage{javascript}{
  keywords={break, case, catch, continue, debugger, default, delete, do, else, finally, for, function, if, in, instanceof, new, return, switch, this, throw, try, typeof, var, void, while, with},
  keywordstyle=\color{keywordblue}\bfseries,
  ndkeywords={class, export, boolean, throw, implements, import, this},
  ndkeywordstyle=\color{keywordblue}\bfseries,
  identifierstyle=\color{black},
  sensitive=false,
  comment=[l]//,
  morecomment=[s]{/*}{*/},
  commentstyle=\color{commentgray}\ttfamily,
  stringstyle=\color{stringred}\ttfamily,
  morestring=[b]',
  morestring=[b]"
}

\unnumbered

\tcbset{
  colback=gray!5!white,
  colframe=gray!75!black,
  boxrule=0.5pt,
  arc=2pt,
  left=6pt,
  right=6pt,
  top=4pt,
  bottom=4pt,
  boxsep=2pt,
  enhanced jigsaw,
  breakable,        
  sharp corners,
  before upper={\ttfamily\footnotesize}, 
}

\tcbset{
  colback=gray!5!white,
  colframe=gray!75!black,
  boxrule=0.5pt,
  arc=2pt,
  left=6pt,
  right=6pt,
  top=4pt,
  bottom=4pt,
  boxsep=2pt,
  enhanced jigsaw,
  breakable,
  sharp corners
}

\newtcblisting{JSCodeBox}{
  listing only,
  listing options={
    language=JavaScript,
    basicstyle=\ttfamily\footnotesize,
    breaklines=true,
    showstringspaces=false,
    tabsize=2
  },
  colback=gray!5!white,
  colframe=gray!75!black,
  title=JavaScript Snippet,
  fonttitle=\bfseries
}

\begin{document}
\title[]{The Impact of Generative AI on Social Media: An Experimental Study}

\author*[1]{\fnm{Anders Giovanni} \sur{M\o{}ller}}\email{agmo@itu.dk}

\author[2,3,4]{\fnm{Daniel M.} \sur{Romero}}\email{drom@umich.edu}
\author[2,4]{\fnm{David} \sur{Jurgens}}\email{jurgens@umich.edu}
\author[1,5]{\fnm{Luca} \sur{Maria Aiello}}\email{luai@itu.dk}

\affil[1]{\orgdiv{Data Science Department}, \orgname{IT University of Copenhagen}, \orgaddress{\street{Rued Langgaards Vej 7}, \city{Copenhagen}, \postcode{2300}, \country{Denmark}}}

\affil[2]{\orgdiv{School of Information}, \orgname{University of Michigan}, \orgaddress{\street{2200 Hayward Street}, \city{Ann Arbor}, \postcode{48109}, \state{Michigan}, \country{USA}}}

\affil[3]{\orgdiv{Center for the Study of Complex Systems}, \orgname{ University of Michigan}, \orgaddress{\street{500 Church Street}, \city{Ann Arbor}, \postcode{48109}, \country{USA}}}

\affil[4]{\orgdiv{Computer Science and Engineering Division}, \orgname{ University of Michigan}, \orgaddress{\street{2260 Hayward Street}, \city{Ann Arbor}, \postcode{48109}, \country{USA}}}

\affil[5]{\orgdiv{Pioneer Centre for AI}, \orgaddress{\street{\O{}ster Voldgade 3}, \city{Copenhagen}, \postcode{1350}, \country{Denmark}}}

\abstract{Generative Artificial Intelligence (AI) tools are increasingly deployed across social media platforms, yet their implications for user behavior and experience remain understudied, particularly regarding two critical dimensions: (1) how AI tools affect the behaviors of content producers in a social media context, and (2) how content generated with AI assistance is perceived by users. To fill this gap, we conduct a controlled experiment with a representative sample of 680 U.S. participants in a realistic social media environment. The participants are randomly assigned to small discussion groups, each consisting of five individuals in one of five distinct experimental conditions: a control group and four treatment groups, each employing a unique AI intervention---chat assistance, conversation starters, feedback on comment drafts, and reply suggestions. Our findings highlight a complex duality: some AI-tools increase user engagement and volume of generated content, but at the same time decrease the perceived quality and authenticity of discussion, and introduce a negative spill-over effect on conversations. Based on our findings, we propose four design principles and recommendations aimed at social media platforms, policymakers, and stakeholders: ensuring transparent disclosure of AI-generated content, designing tools with user-focused personalization, incorporating context-sensitivity to account for both topic and user intent, and prioritizing intuitive user interfaces. These principles aim to guide an ethical and effective integration of generative AI into social media.}

\keywords{Generative Artificial Intelligence, Human-Computer Interaction, Controlled Experiment, Large Language Models}



\maketitle

\section{Introduction}\label{sec1}

The rapid integration of artificial intelligence (AI)-driven text generation tools into social media platforms is reshaping how users create and engage with content, raising new questions about their effects on the quality and dynamics of online interactions~\cite{Ziems_Held_Shaikh_Chen_Zhang_Yang_2024, Xi_Chen_Guo_He_Ding_Hong_Zhang_Wang_Jin_Zhou_etal._2025}. AI writing tools notably reduce barriers to content creation by lowering required effort and expertise. Although these technologies are increasingly adopted across sectors such as journalism~\cite{Anantrasirichai_Bull_2022, Pavlik_2023}, education~\cite{Yan_Greiff_Teuber_Gasevic_2024}, and creative industries~\cite{Anantrasirichai_Bull_2022}, their potential impact is particularly pronounced in social media contexts. Social media platforms play a central role in shaping public discourse, influencing democratic engagement, and enabling rapid, large-scale dissemination of information and ideas~\cite{Jennings15032021}. Therefore, introducing AI into these platforms may reshape dynamics of interaction, authenticity, and nature of online discussions~\cite{Jakesch_Bhat_Buschek_Zalmanson_Naaman_2023, Bail_2024}. Previous studies highlight the promise of AI assistance in advancing human creativity~\cite{Wingstrom_and_Lundman_2024}, increasing user engagement~\cite{Rattanasevee_Akarapattananukul_Chirawut_2024}, and facilitating broader inclusivity in online discussions~\cite{Mikhaylovskaya_2024, Tessler_Bakker_Jarrett_Sheahan_Chadwick_Koster_Evans_Campbell-Gillingham_Collins_Parkes_et_al_2024}. Yet, this optimism is tempered by concerns about potential drawbacks, such as declining content quality~\cite{ai_content_quality}, proliferation of misinformation~\cite{Drolsbach_Solovev_Prollochs_2024}, and diminished authenticity of user interactions~\cite{Osborne_Bailey_2025}.

Empirical evidence quantifying how AI assistance reshapes online participation dynamics, content quality, and user perceptions remains scarce, particularly in realistic, platform-integrated scenarios~\cite{Yang_Singh_Menczer_2024}. Addressing this critical gap, our study provides empirical insights through a controlled experiment conducted on a realistic social media platform. We approach the debate about AI-assisted content creation from two complementary perspectives: first, how AI assistants in social media affect the experience of content \textit{producers}, and second, how these interventions shape \textit{consumers}' perceptions of content. Our experiment fills a crucial gap by directly assessing how AI assistants transform both producer and consumer experiences on social media. 

We conduct the experiment using a custom-built platform that closely simulates an online chat room resembling discussions on common social media forums. A representative sample of 680 U.S. participants is partitioned into groups of five people, who are then randomly assigned to one of five conditions: one control (no AI assistance) and four non-overlapping treatment conditions, each featuring distinct AI interventions previously proposed for enhancing online interactions. These interventions include (1) an open-ended \textit{chat} with an AI assistant~\cite{argyle_leveraging_2023}, (2) AI-generated reply \textit{suggestions} with varying stances (agreeing, neutral, disagreeing)~\cite{Jakesch_Bhat_Buschek_Zalmanson_Naaman_2023,Di_Fede_Rocchesso_Dow_Andolina_2022}, (3) AI-driven \textit{feedback} on comment drafts~\cite{Yan_Greiff_Teuber_Gasevic_2024, ziegenbein_llm-based_2024}, and (4) AI-generated \textit{conversation starters}~\cite{Do_Kong_Lee_Bailey_2022}; interventions are described in more detail in the Methods section and in ``AI Prompts and Settings" in the \textit{Supplementary Information} (see Fig.~\ref{fig:platform-ui} for platform screenshots). Together, these interventions represent a comprehensive range of AI-based approaches considered by both researchers and industry practitioners for enhancing online interactions. To assess the heterogeneity of the interventions' effects across topics, participants sequentially discuss three randomly-ordered topics---ranging from conversational (\textit{cats vs. dogs}), to scientific (health benefits of \textit{oats}), to political (\textit{universal basic income})---with each topic limited to a 10-minute interaction. We assess various proxies for user engagement and quality of experience through questionnaires before and after participation, and further track participants' interactions on the platform---including comments, reactions, and AI usage---to comprehensively evaluate how AI interventions impact both content producers and consumers. 

\begin{figure}[htbp]
  \centering
  \includegraphics[width=0.97\textwidth]{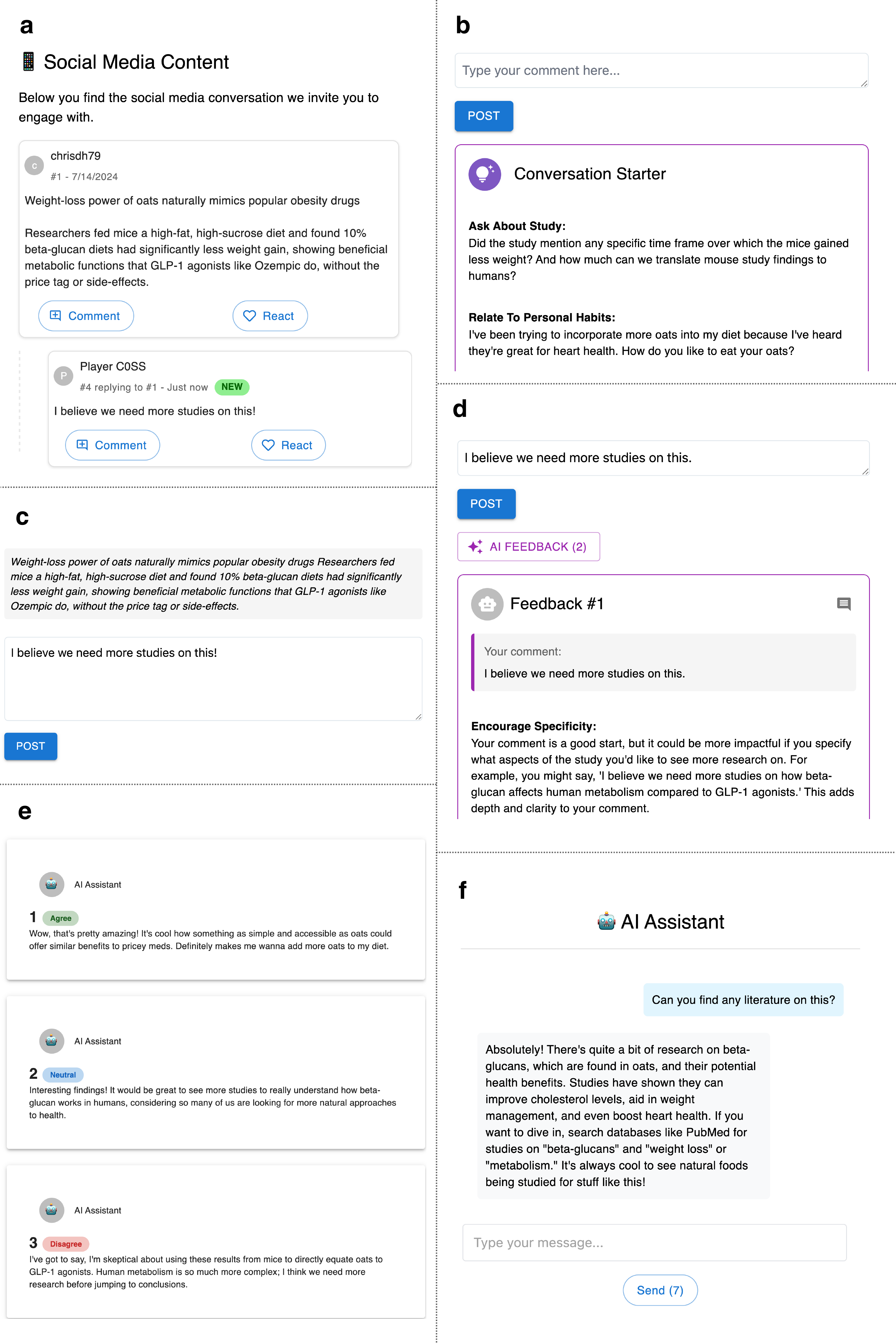}
  \caption{\textbf{User interface of the experimental platform.} 
  \textbf{a}, Main discussion thread. \textbf{b}, Conversation Starter. \textbf{c}, Comment modal. \textbf{d}, Feedback tool. \textbf{e}, Suggestions tool. \textbf{e}, Chat assistant.}
  \label{fig:platform-ui}
\end{figure}


\begin{figure*}
\centering
\includegraphics[width=0.80\textwidth]{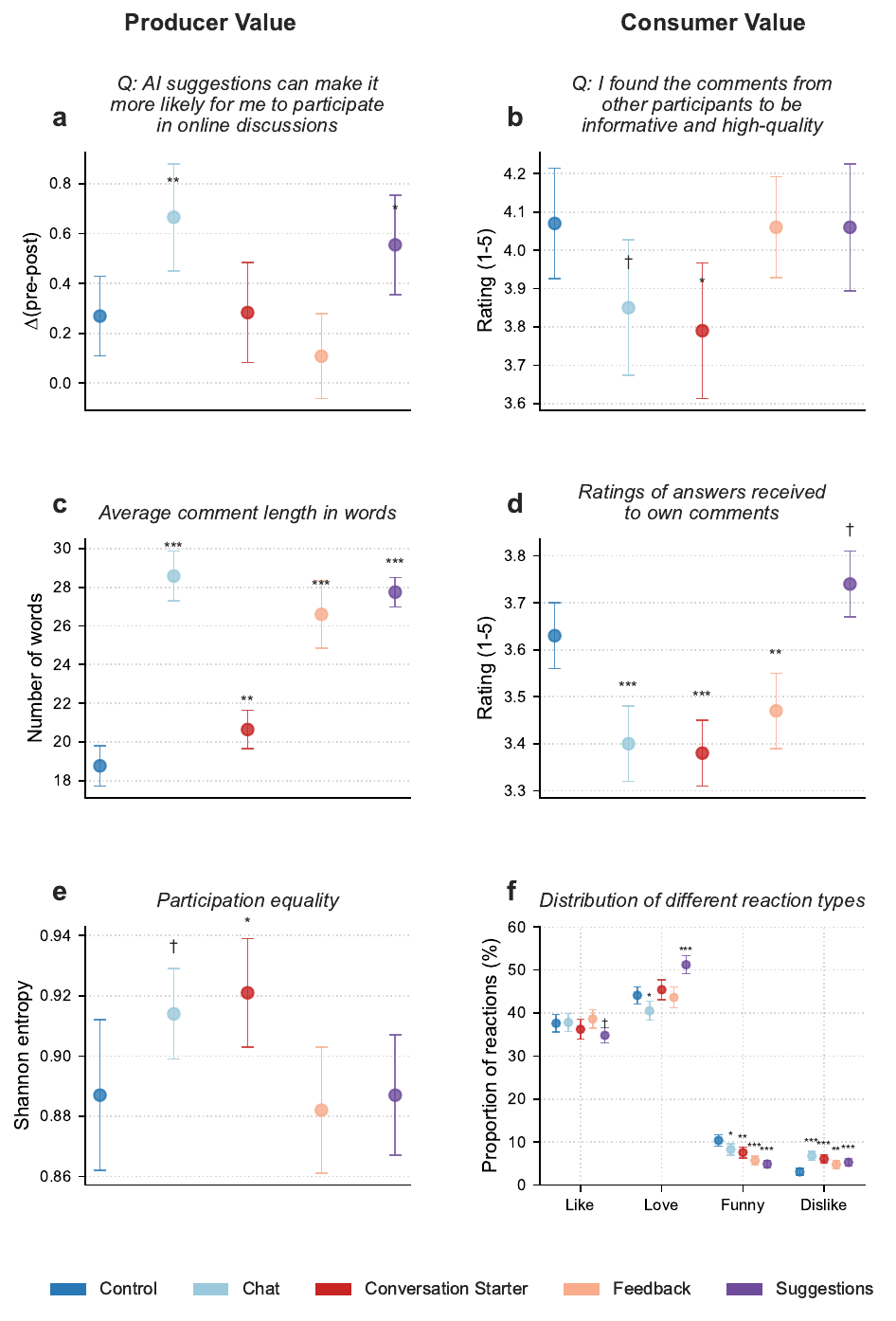}
\caption{
    \textbf{Producer value} results (left column) and \textbf{consumer value} results (right column). \textbf{a}, Change in responses to ``\textit{AI suggestions can make it more likely for me to participate in online discussions}" (1-5 Likert scale) before and after the study. \textbf{b}, Likert-scale responses to ``\textit{I found the comments from other participants to be informative and high-quality}". \textbf{c}, Average comment length in words. \textbf{d}, Individual user ratings of replies received on their own comments. \textbf{e}, Participation equality measured using normalized Shannon entropy based on participation distribution per round. \textbf{f}, Distribution of reaction types (Like, Love, Funny, Dislike) across conditions. \textdagger\textit{p} $<$ 0.10; *\textit{p} $<$ 0.05; **\textit{p} $< 0.01$; ***\textit{p} $<$ 0.001. For details on statistical tests and bootstrapping procedures, see \textit{Supplementary Information}.
    }
\label{fig:figure-1}
\end{figure*}

Overall, our findings indicate that AI assistance substantially influences both user-generated content quality and consumer perception, although with notable variation across interventions. AI-supported participants demonstrate increased engagement and content production metrics, but these improvements are associated with nuanced, sometimes negative, shifts in consumer perceptions and reactions. Critically, no single AI tool enhances both producer and consumer experiences, highlighting complex trade-offs. 

AI interventions generally increase participants' willingness to engage and improve aspects of content creation from the producer perspective. Participants using the \textit{Chat} and \textit{Suggestions} features notably report that the AI would increase their willingness to participate in online discussions, compared to control (Fig.~\ref{fig:figure-1}a). Additionally, all AI-supported tools significantly increase the average length of user comments (Fig.~\ref{fig:figure-1}c). Participation equality among users, measured by normalized Shannon entropy based on the proportion of comments per user in each round, noticeably improves under the \textit{Conversation Starter} intervention, indicating more balanced participation (Fig.~\ref{fig:figure-1}e). Regression analysis for each treatment condition indicates that only \textit{Conversation Starter} significantly increased the likelihood of receiving a reply, while the other AI-tools showed positive but non-significant signals (see \textit{Supplementary Information}, ``Post-hoc Regression Analyses on Treatment Impact of Reply Likelihood").

From the consumer perspective, none of the AI interventions improve user perceptions compared to the control condition. All AI treatments significantly increase `Dislikes', while \textit{Suggestions} evoke more `Love' reactions (Fig.~\ref{fig:figure-1}f). Participants perceive comments to be less informative and lower quality in both the \textit{Chat} and \textit{Conversation Starter} conditions compared to control (Fig.~\ref{fig:figure-1}b). Similarly, users rated replies to own comments significantly lower in all but the \textit{Suggestions} condition, which uniquely show positive, though weakly significant, ratings compared to control (Fig.~\ref{fig:figure-1}d).

Together, these findings highlight a critical duality: although AI interventions broaden participation and increase the volume of content, they also risk creating ``semantic garbage" perceived as lower quality than human-generated text~\cite{floridi2020gpt} and degrading the quality of subsequent human conversation once AI is introduced to a thread. Our experiment stands as one of the first direct tests of whether AI can genuinely elevate online discussions or merely amplify low-quality interaction in a way that clouds its potential benefits---an outcome our evidence suggests is more likely. In exploratory regression analyses assessing whether demographics (age, gender, sex, education, and political affiliation) would alter these results, we observe only minor effects that did not reach statistical significance. While preliminary, this likely reflects limited sample sizes across treatment groups (see \textit{Supplementary Information} for additional details).

\section{How are the AI tools used?}

\begin{figure*}[th!]
\centering
\includegraphics[width=0.94\textwidth]{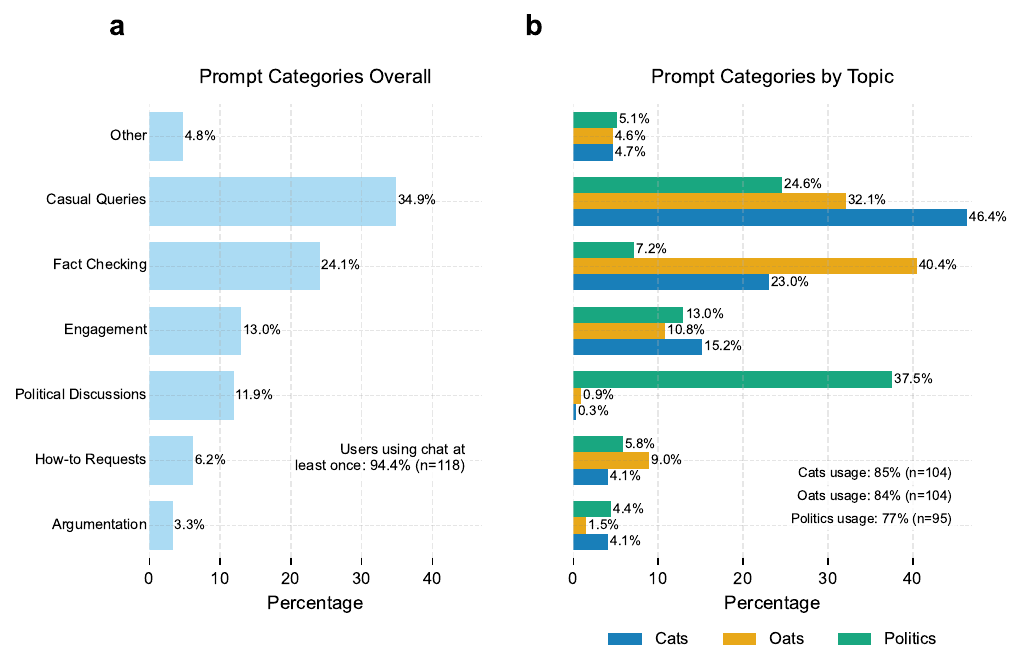}
\includegraphics[width=0.94\textwidth]{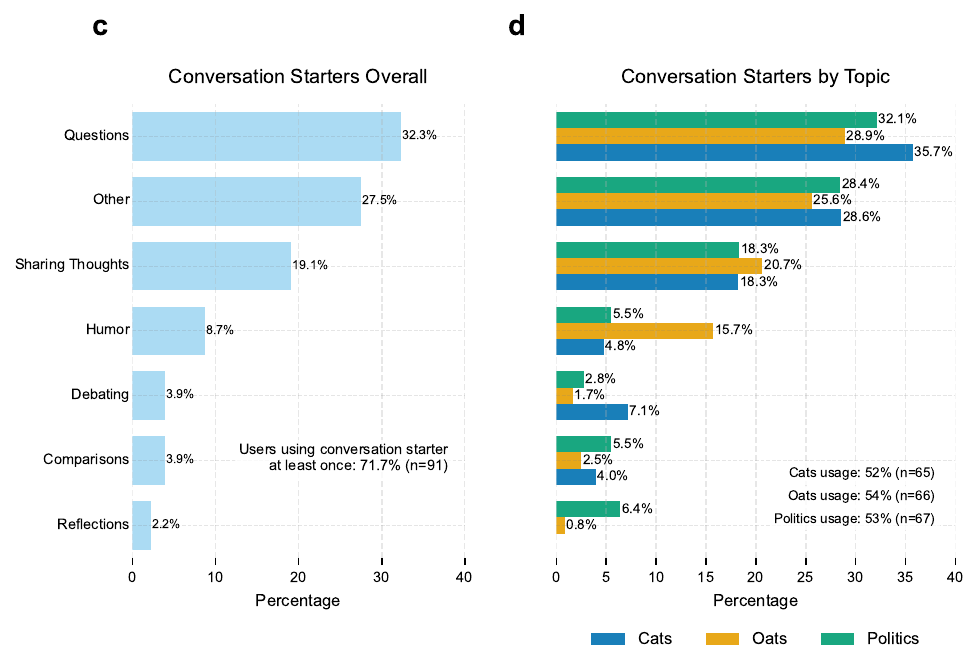}
\caption{\textbf{Treatment usage}. \textit{Chat}: \textbf{a}, Distribution of prompt categories overall. \textbf{b}, Prompt categories by topics. \textit{Conversation Starter}: \textbf{c}, Distribution of conversation starters overall. \textbf{d}, Conversation starters by topic. Usages values indicate the fraction and number of participants who used the AI tool within topics. Overall proportions below 1\% are excluded.} 
\label{fig:chat-conversation}
\end{figure*}

\begin{figure*}[th!]
\centering
\includegraphics[width=0.87\textwidth]{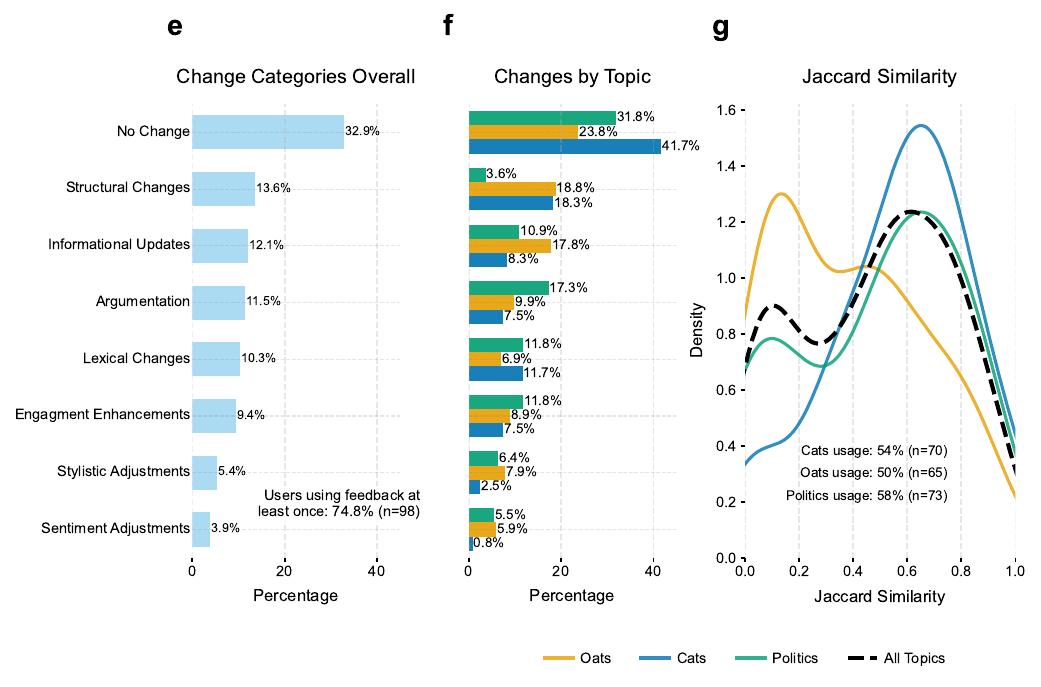}
\includegraphics[width=0.87\textwidth]{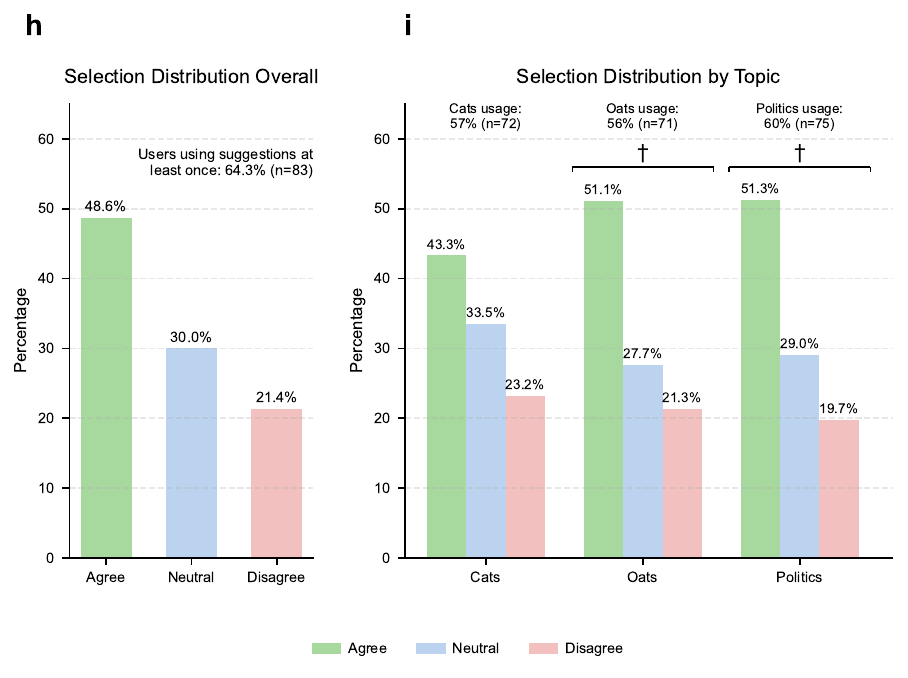}
\caption{
\textbf{Treatment usage}. \textit{Feedback}: \textbf{e}, Distribution of feedback changes overall. \textbf{f}, Feedback changes by topic. \textbf{g}, Distribution of Jaccard similarity scores between original and revised comments overall and by topic. \textit{Suggestions}: \textbf{h}, Overall distribution of selected suggestions. \textbf{i}, Selected suggestions by topic. Significance markers are based on Chi-square tests of independence comparing selection distributions between \textit{cats} and \textit{oats}, and \textit{cats} and \textit{politics}: \textdagger\textit{p} $<$ 0.10; *\textit{p} $<$ 0.05; **\textit{p} $<$ 0.01; ***\textit{p} $<$ 0.001.
}
\label{fig:feedback-suggestions}
\end{figure*}

Building on the duality---AI interventions boosting participation yet risking polluting conversation with low-quality content---we also uncover nuanced differences in how participants employed each tool. Usage patterns are far from uniform and often reflect the tools intended design and conversational context. These variations merit closer auditing, both to clarify which AI-aid paradigm holds the most promise for meaningful engagement and to guide future employments and refinements. Figs.~\ref{fig:chat-conversation}a-d and~\ref{fig:feedback-suggestions}h-i detail key usage patterns (see the \textit{AI Usage Analysis} section and ``AI Usage" in \textit{Supplementary Information} for details on the analyses).

\subsection{Chat provides flexibility but engagement varies across context}

\noindent
For the \textit{Chat} feature, we find widespread adoption, with $94.4\%$ ($n=118$) of treatment participants engaging at least once with the AI, resulting in a total of $960$ user prompts (Fig.~\ref{fig:chat-conversation}a,b). Users predominantly interacted with the AI for topic-related, informal, exploratory dialogue (\textit{Casual Queries}, $34.9\%$, $n=335$)---for instance, in the \textit{cats} discussion: `One of my cats getting stuck in a jar'. However, a substantial proportion of queries was used for \textit{Fact Checking} ($24.1\%$, $n=231$), highlighting the AI's role as an informational resource. Additionally, smaller proportions of interactions involved \textit{Engagement} ($13.0\%$, $n=125$), referring to instances where users sought assistance with crafting responses. A similar share of prompts involved \textit{Political Discussions} subjects ($11.9\%$, $n=114$), reflecting users interest in issues with political or societal implications. Disaggregating the usage patterns by topical context, the nature of the engagement with the AI strongly vary: casual queries dominated the lighter subject of \textit{cats} ($46.4\%, n=159$), whereas participants primarily used the AI for fact checking in the scientific conversation of \textit{oats} ($40.4\%$, $n=131$), and for political discussions in the divisive topic of \textit{politics} ($37.5\%$, $n=110$). This adaptability suggests that participants tailored their AI-usage to the tone of each conversation. The features' flexibility and the users' likely familiarity with the interface contributed to the observed increases in engagement and production metrics. Yet, this may have induced verbose contributions, as reflected in longer sentence lengths (\textit{Chat}: 28.59 words, vs \textit{Control}: 18.76 words), which could explain the relatively low perceived quality and informativeness of comments, compared to the control condition (see Fig.~\ref{fig:figure-1}b). 

\subsection{Conversation Starters lower barriers to entry but are often misaligned with user intent}
The \textit{Conversation Starter} feature, designed to spark initial engagement or enrich ongoing interactions, was used by $71.7\%$ ($n=91$) of participants at least once, resulting in a total of $345$ uses (Fig.~\ref{fig:chat-conversation}c,d). When applying these starter suggestions, participants predominantly used open-ended or exploratory hints (\textit{Questions}, $32.3\%$, $n=115$). Yet a notable share of user comments ($27.5\%, n=98$) diverged from the AI-generated conversation starters entirely, suggesting users often dismiss the AI recommendations. This pattern of selective adoption was consistent across topics---usage rates around 52-54\% for all subjects---reflecting a stable preference for exploratory, curiosity-driven engagement regardless of the conversational context. Although the \textit{Conversation Starter} did help lower barriers to participation, user behavior and questionnaire answers suggest the aid regularly misaligned with their communication goals. 

\subsection{Feedback refines arguments in high-stakes discussions but is ignored when stakes were low}

Participants' integration of AI-generated \textit{Feedback} varied substantially by topic, reflecting diverse user priorities across context (Fig.~\ref{fig:feedback-suggestions}e-g). In the casual \textit{cats} discussion, users typically made minimal or no textual changes after receiving AI feedback, with 41.7\% of cases ($n=50$) showing no edits to the original comment before submitting. By contrast, in the scientifically grounded \textit{oats} discussion, users more often revised their comments, often incorporating \textit{Structural Changes} ($18.8\%, n=22$) or making \textit{Informational Updates} ($17.8\%, n=18$). In \textit{politics}, the most common revisions involved \textit{Argumentation} ($17.3\%, n=19$), highlighting the AI's role in supporting debate-oriented discourse. Overall adoption of the feedback feature was high ($74.8\%$, $n=98$), with a total of $331$ uses, although per-topic usage ranged between $49.6\%$ and $57.9\%$ of the participants. These patterns indicate that users engaged with the \textit{Feedback} tool more frequently on contexts where credibility, persuasion and clarity seemed most important, reflecting an intrinsic motivation for rhetorical strength.

\subsection{In divisive contexts, users prefer AI suggestions that express agreement}

The \textit{Suggestions} feature, which offers context-dependent responses across agreeing, neutral, or disagreeing stances, showed moderate adoption ($64.3\%$, $n=83$), with a total of $1,197$ generated suggestions selected (Fig.~\ref{fig:feedback-suggestions}h,i). Overall, participants predominantly selected agreeing suggestions ($48.6\%, n=582$), with neutral ($30.0\%$, $n=359$) and disagreeing ($21.4\%$, $n=256$) responses selected less often. Users' preference for agreement varied by topic sensitivity: participants favored agreeing responses more strongly in higher-stakes or sensitive topics (\textit{oats}: $51.1\%$, \textit{politics}: $51.3\%$) compared to the less contentious topic (\textit{cats}: $43.3\%$). This pattern suggests a tendency to avoid divisive positions on controversial discussions, whereas disagreeing stances were more acceptable in lighter, low-risk conversations. 

\subsection{Participants value supportive AI but want more personalization}

When asking participants for open-ended feedback on the AI tools after the study ($n=245$ participants submitted written responses in the optional feedback field), users consistently emphasized the value of AI assistance for generating ideas, clarification, and initiating engagement---in particular when they felt ``lost for words.'' Despite differences across tools, users frequently described the AI as helpful for fact checking, comment reflection, and simplifying the process of contributing to the conversations. The tools would also help keep interactions constructive. In contrast, participants felt the AI content lacked authenticity and came across as overly generic, and requested more personalization of the AI---such as adapting to users' style and context. Users reported that the \textit{Chat} feature was useful for fact checking or exploring unfamiliar topics, noting that it made them feel more informed and confident in joining discussions. Still, some found the responses unnaturally formal or robotic. The \textit{Conversation Starter} was valued for inspiration, especially for posing questions, but was also described as impersonal. The \textit{Feedback} feature helped users reflect more deeply on their comments, though some noted that this could become exhausting or overly analytical. For \textit{Suggestions}, users appreciated the simplicity and effectiveness in enabling quick replies. This made engagement more accessible, but some users found the suggestions impersonal.

\noindent

\section{How to move forward}

These results entail important implications for the deployment of generative AI on social media platforms. Although participants in the treatment conditions produced more content, they often recognized it as generic, impersonal, and of lower quality. Such reduced perceived authenticity diminishes trust and informational value---risks that align with recent concerns about AI-assisted content flooding digital platforms with low-quality text~\cite{Feuerriegel_DiResta_Goldstein_Kumar_Lorenz-Spreen_Tomz_Prollochs_2023, Yang_Menczer_2024, Wei_Tyson_2024}. A central concern is that widespread use of generative AI may saturate platforms with superficial or generic content, diluting the visibility and impact of more original contributions---and, as our findings suggest, lowering the quality of subsequent conversations within threads, even among users not using the AI themselves. 

Nevertheless, our findings show conditions under which AI tools may enhance public discourse. Participants valued the AI for idea generation, clarity, and initiating engagement---particularly in perceived high-stake or cognitive demanding topics. If tools are developed with greater personalization, contextual awareness, and stylistic nuances, they could support more inclusive and constructive interactions---particularly for individuals typically hesitant to engage in public discussions. To guide ethical and effective deployment, we propose four design principles:

\begin{enumerate}
    \item \textbf{Appeal of optional use and transparent disclosure}: 
    Our findings suggest that AI tools are most positively received when offered as optional. Across all treatment conditions, usage patterns indicate selective engagement. In the \textit{Chat} condition, participants submitted an average of 3.36 to 3.55 prompts per round, with only 13-16\% of their final comments showed direct textual overlap. This suggests that users primarily engaged with the tool for ideation, clarification, or rhetorical guidance rather than direct copying. Similarly, for the \textit{Conversation Starter}, only between 16\% and 19\% of submitted comments overlapped with AI suggestions---indicating that users often adapted, modified, or disregarded them based on text quality or alignment with their intent. The \textit{Feedback} tool, although more cognitively demanding and less widely adopted, proved highly effective when used:  in over 89\% of cases where participants used the tool when composing a comment, they subsequently submitted it to the discussion.

    These behavioral signals are mirrored by open-ended questionnaire responses. Participants reported appreciating AI support for generating ideas, retrieve information, and overcoming initial writer's block. Most treatment participants indicate they would use such tools if integrated into existing social media platforms (mean ratings on a 1-5 Likert scale: \textit{Chat}: 3.97, \textit{Conversation Starter}: 3.54, \textit{Feedback}: 3.82, \textit{Suggestions}: 3.94). At the same time, users noted easy identification of AI-generated content in others' posts---often describing the text as robotic, generic, or impersonal in the open-ended feedback. This perception was also found in the exit questionnaire: the proportion of participants believed by the other users to use AI tools ranged from from 13.8\% in the control group to 37.7-43.6\% across treatments (Control: 13.8\%, \textit{Chat}: 38.9\%, \textit{Conversation Starter}: 37.7\%, \textit{Feedback}: 39.1\%, \textit{Suggestions}: 43.6\%).
    This perception of diminished authenticity highlights a core challenge: although AI can reduce cognitive barriers and enhance participation, overuse may erode trust in the overall discourse.

    Together, these insights motivate a core design principle: transparent disclosure paired with user autonomy. In particular, content directly copied from generative AI tools should be clearly labeled~\cite{Gamage_Sewwandi_Zhang_Bandara_2025, Gallegos_Shani_Shi_Bianchi_Gainsburg_Jurafsky_Willer_2025}, as such use might affect perceptions of authenticity. However, more nuanced uses---such as revising a draft with AI assistance or drawing inspiration from AI suggestions---may not require explicit labeling. Platform policies should account for these distinctions and support user agency in deciding how and when to engage with AI content. With this, platforms can preserve perceptions of authenticity~\cite{Kirkby_Baumgarth_Henseler_2023}, improve trust in human-AI interaction, and accommodate diverse user preferences.

    \item \textbf{Personalization}: Across conditions, participants reported a desire for greater personalization of AI-generated content, more closely reflecting their individual voice, tone, and communicative intent. In the open-ended feedback, 34 participants explicitly described the AI responses as overly generic, noting it often felt impersonal.
    These perceptions are reflected in participants' ratings of comment quality and informativeness, as well as their assessments of replies received to their own comments, with treatment groups rated lower than the control~(Fig.~\ref{fig:figure-1}b,d). This suggests that although AI can support content generation, failing to adapt to user-specific context and style may lower perceived value of the interaction.
    Notably, in the \textit{Feedback} condition---where users are guided in refining their own comments---we observe an increase in perceived value alignment with other participants, an increment from 33.9\% to 42.5\% (see ``Ratings of Users and Comments" in \textit{Supplementary Information}). This may indicate that tools enabling reflective personalization, rather than generic generation, could cultivate better identification and resonance among users.

    These behavioral and perceptual patterns underline the importance of AI systems that adapt to individual users. Future tools should incorporate stylistic adaptation and learn from prior interactions to provide outputs that is perceived personal, relevant, and authentic. 

    \item \textbf{Contextual awareness, flexibility, and authenticity}: Effective AI support in social media conversations requires responsiveness to the context of the discussion. In our study, participants engaged with the AI in varying ways depending on topic sensitivity. 
    
    In the \textit{Chat} condition, users most often engaged the AI with casual, topic-related prompts during the \textit{cats vs. dogs} discussion---asking open-ended questions that reflected a conversational, informal, and low-effort use of the tool. In the scientific \textit{oats} discussion, participants frequently used the AI as an informational resource, asking for facts and clarification. In contrast, political discussion queries were most common in the political topic, where users asked for reflections, viewpoints, and pros and cons---using the AI more analytically to explore perspectives and argumentative reasoning. This adaptive usage suggests that participants naturally attuned expectations for AI based on the conversational subject. In the \textit{Suggestions} condition, users preferred agreement responses in the higher-stakes \textit{oats} and \textit{politics} discussions, but were more willing to express disagreement or neutrality in the \textit{cats} topic. In the \textit{Chat} condition, 12 out of 58 participants who provided open-ended AI-feedback specifically noted that the AI tool was helpful when engaging with unfamiliar topics---supporting that AI can lower barriers to entry in online discussions. However, participants reported lack of personal style and authenticity, mismatching individual demands in the conversation. This points to a central design challenge: fixed-responses that ignore topical and personal risk undermining authenticity and usefulness of AI-assisted content. 

    These findings motivate a design principle for AI tools to embed and understand contextual sensitivity. This imply adjusting conversational intent, adjusting tone, stance, and formality based on topic domain and personal preferences. Informality may help in trivial discussions, but scientific or political contexts appeal toward factual elements, rhetorical nuances, and societal understanding. Systems that fail to accommodate such topical diversity may reduce trust or suppress meaningful engagement. Flexibility and authenticity---grounded in both topic and user intent---should be central to the development of socially integrated AI systems on social media.
    
    \item \textbf{Familiar and user-friendly interfaces}:
    Ease of use is central in the adoption and effectiveness of AI tools for content creation on social media platforms. In the \textit{Chat} condition, designed with a familiar and low-friction sidebar interface, we find the most wide adoption (94.4\% used it at least once). In contrast, the \textit{Feedback}, \textit{Conversation Starter}, and \textit{Suggestions} tools, showed lower adoption rates (\textit{Feedback}: 74.81\%, \textit{Conversation Starter}: 71.65\%, \textit{Suggestions}: 64.34\%). This underlines the trade-offs between tool abundance and user effort. These behavioral and perception perspectives affirm the importance of intuitive and convenient interfaces in encouraging tool engagement. Participants in the \textit{Chat} and \textit{Suggestions} conditions find these tools as more supportive overall---reporting that they made participation easier, felt more intuitive to use, and led to higher-quality contributions---compared to those using \textit{Feedback} and \textit{Conversation Starter} (see ``Supplementary Figures" in \textit{Supplementary Information}).

    Altogether, these findings support a key design principle: simplicity and familiarity in user interfaces are vital to enhance adoption and usage. When AI tools are embedded seamlessly into platform workflows---with easy entry points and low interaction costs---they are more regarded as a supportive mechanism rather than disruptive. Future implementations should prioritize clarity and accessibility by embedding AI tools into the natural flow of conversation, minimizing user effort without sacrificing optional depth.
\end{enumerate}

To more deeply understand the dynamics of AI-assisted interactions, future work should build on this experimental foundation by scaling across time, platforms, and population. One critical extension involves capturing temporal AI tool usage---how engagement patterns change over longer exposure, whether the innovation effects diminish, and how personalization adapt over longer interactions. Expanding the sample size would also support more robust cross-treatment comparisons and enable better analysis of subgroup diversity. Deploying similar interventions on other social platforms---within controlled environments---would offer deeper ecological validity and understanding of how AI tools affect different social ecosystems online. Our study establishes a baseline for controlled experimentation with integrated AI support for content-creation, but continuous progress requires research across diverse online environments.

An ethical deployment of AI tools on social media necessitates continuous auditing. This includes not only how user data is processed and how model biases are mitigated, but also how such tools may affect and reshape collective behavior over time. Additionally, transparency in AI usage, guardrails to prevent misinformation and marginalization, and mechanisms for user control must be built into deployments~\cite{Drolsbach_Solovev_Prollochs_2024}. As AI tools are introduced into social media environments already designed optimized for sustained engagement, they may further influence conversational dynamics. Our controlled experiment captures short-term, user-level interaction dynamics. Long-term effects of AI assistance on social media will emerge through cumulative patterns of use, repeated exposure, and feedback loops. Even subtle interventions---such as nudging users toward agreement or increasing comment length---may gradually change conversational norms, influence tone, speed of engagement, and the inclusiveness of discussions. Societal impact remains uncertain, but AI tools can alter how persuasive, sensitive, or political content disseminates, with potentially impactful consequences for the fragile integrity of public discourse. 

AI content-generation tools are naturally becoming a fundamental layer of digital interaction. Our findings highlight both promise and risks. When designed with transparency, personalization, and contextual flexibility, these tools can lower cognitive barriers, broaden participation, and support more inclusive engagement. But if deployed without careful consideration of their emergent effects, social media platforms risk saturating public discourse with generic, inauthentic content, undermining quality and trustworthiness of online conversations. The future of democratic communication online will depend not only on the capabilities of AI tools, but on how thoughtfully they are embedded into the social communicative ecosystem of digital platforms.

\section{Methods}

In our experiment, participants progressed through a structured experimental flow. After providing informed consent, they (i) completed a pre-study questionnaire assessing demographics and attitudes toward social media and AI. Participants were then (ii) onboarded to the platform and the task via instructions and a demonstration video, (iii) engaged in three 10-minute discussions in randomized order, (iv) completed a post-study questionnaire evaluating their experience and rating content and other participants, and (v) finally received their compensation token. Full experimental details, questionnaire items, distribution of answers, seed content, and evaluation are available in the \textit{Supplementary Information}. 

Our experimental setup complies with relevant ethical regulations. University of Michigan Institutional Review Board (IRB) approved the study protocol (HUM00258995). Participants were compensated \$9.50 USD for an estimated 45-minute session, corresponding to an hourly rate of \$12.66.

\subsection{Platform Design}

We built a custom online discussion platform to simulate a typical forum-based social media environment, adopted into Empirica~\cite{Almaatouq_Becker_Houghton_Paton_Watts_Whiting_2021} to handle experiment logistics (see Fig.~\ref{fig:platform-ui} for platform screenshots). The design was inspired by platforms like Reddit with threaded conversations and lightweight user interaction. Participants could post comments and react using seven emoji-based responses (like, love, dislike, angry, wow, sad, funny), with ``love" set as the default. A \emph{U.S.-representative} sample of 680 participants---recruited via Prolific between January 3 and January 21, 2025---was randomly put into groups of five to mimic small-group online discussions and maintain experimental control\footnote{Prolific optionally supports recruitment of U.S.-representative samples based on key demographics such as age, sex, ethnicity, and political affiliation. See~\url{https://researcher-help.prolific.com/en/article/e6555f}}. Each group was randomly assigned to one of the five experimental conditions: a control group with no AI assistance, or one of the four AI-assisted treatment conditions. Participants interacted only within their assigned group and had no prior knowledge about the other users. Each group engaged in three 10-minute discussions on different predefined topics on varying social sensitivity and complexity: one trivial (\textit{cats vs. dogs}), one scientific (health benefits of \textit{oats}), and one political (\textit{universal basic income}). To control for order effects, the sequence of topics was randomized across groups. We source the initial seed data---the content shown at the start of each 10-minute discussion---from existing Reddit threads~\cite{reddit_topic_catsdogs, reddit_topic_oats, reddit_topic_ubi}. Each comment displayed a timestamp, interactive buttons for commenting and reacting, visual indentation to indicate thread structure, and a parent ID referring to the comment it responded to. Reactions would be shown to all participants but without identification of who reacted. New comments were highlighted with a brief alert and a ``new" tag lasting one minute to support real-time flow.

The platform was designed to minimize distractions and reduce cognitive load, allowing participants to focus on the conversation. The setup ensured controlled conditions while preserving key elements of real-world online discussions. Before entering the discussion platform, participants were shown a brief onboarding interface introducing the platform and their task. This included a written description of the discussion structure and a short demonstration video showing how to navigate the platform, post comments, and react to content. For participants assigned to AI-assisted conditions, additional instructions were provided explaining the specific tool available. The demonstration video included a walkthrough of how to access and use the AI feature. During each 10-minute discussion round, we log user activity at the individual level, including comments posted, reactions given, and use of AI-tools.

\subsection{AI Tools}
To test the impact of different paradigms of AI assistance in social media discussions, we integrated four distinct AI tools into the platform. Each tool was designed to reflect approaches to AI-assisted communication found in academic literature and commercial products. All tools were powered by GPT-4o and with a custom prompt tailored to each intervention (see ``AI Prompts and Settings" in \textit{Supplementary Information}). Participants were not informed which AI model was used in treatments or  whether any content they encountered was AI-generated.

\bmhead{Chat}
The \textit{Chat} tool allowed for open-ended interaction with an AI assistant through a sidebar window displayed alongside the conversation thread (Fig.~\ref{fig:platform-ui}f). Participants could engage with the assistant up to eight times per discussion topic. The interface supported informal querying, idea generation, and clarification, giving users flexibility to steer the interaction as desired.

\bmhead{Conversation Starter}
The \textit{Conversation Starter} generated AI-suggested openings for participation, aimed at lowering barriers to entry and stimulating discussion (Fig.~\ref{fig:platform-ui}b). The tool was accessed through a separate button on each comment. The conversation starting suggestions were context dependent but could include follow-up questions, engaging comments, and reflective or contextual statements.

\bmhead{Feedback}
The \textit{Feedback} tool offered real-time guidance on draft comments prior to submission (Fig.~\ref{fig:platform-ui}d). Once users began typing a comment, they could click on a dedicated ``AI Feedback" button to receive tailored suggestions on how to improve their comment. The feedback varied based on context and comment draft, but could include ideas to clarifying arguments, add personal anecdotes, or maintain a balanced tone. The feedback appeared inline below the draft comment, and users could receive three rounds of feedback per comment.

\bmhead{Suggestions}
The \textit{Suggestions} feature provided three AI-generated replies---each adopting a distinct stance (agree, neutral, disagree)---in response to any selected comment (Fig.~\ref{fig:platform-ui}e). The tool was accessed through the comment modal. Participants could regenerate a new set of suggestions up to three times per comment, allowing them to explore alternatives before selecting a reply.\\

\noindent
Each of these tools was embedded into the platform interface to mirror familiar social media interactions while maintaining clarity and minimalism. The tools were designed to be optional, harmoniously integrated into the platform, and supportive of the natural flow of the discussion.  

\subsection{Questionnaires}

\bmhead{Pre-study Questionnaire}
The pre-study questionnaire gathered demographic information (age, gender, education level, occupation, and political affiliation) and attitudes toward online discourse and AI. Participants reported their typical engagement with social media, perceived quality of online discussions, trust in user-generated content, and perceived barriers to participation. Responses were recorded using multiple-choice items and five-point Likert scales. Participants were also asked about their attitudes toward the use of AI on social media platforms. This included perceived impact of AI on participation, comfort, content quality, misinformation, toxicity, polarization, and the need for regulation. These questions also used Likert-scale responses. The full set of questions is provided in \textit{Supplementary Information} under ``Pre-study Questionnaire".

\bmhead{Post-study Questionnaire} Following the experiment, participants answered a second questionnaire capturing their experience on the platform. This included questions about how their participation compared to typical online behavior, their perceptions of quality, and trust in other users. Participants in AI treatment conditions were asked to evaluate the usefulness, quality, and relevance of the AI tool they used. Users were also invited to provide open-ended feedback on their experience with the AI and the platform. To assess perceptual changes, participants answered the same questions about AI on social media from the pre-study questionnaire. This allowed us to quantify shifts in attitude as a result of the experiment. Finally, participants were asked to evaluate the quality of social interaction in their group. They rated 10 replies received on their own comments using a five-point scale to assess the value in the discussion. They also evaluated the other participants in their group on perceived politeness, engagement, political agreement, shared values, use of AI, and whether they might be a bot. Full set of questionnaire items is provided in the \textit{Supplementary Information} under ``Post-study Questionnaire".

\subsection{Evaluation}

We evaluated both behavioral engagement and perceptual effects across producer and consumer perspectives, using a combination of tracked user behavior and self-reported measures from the questionnaires. We used Shannon entropy to measure participation equality within each group, based on the proportions of comments made by each user in a given round. Details on this calculation are provided in \textit{Supplementary Information} under ``Participation Equality (Normalized Shannon Entropy)". We also tracked AI tool usage, allowing us to assess adoption patterns. To evaluate the perceived quality of the interactions, we analyzed responses from the post-study questionnaire. 

For all behavioral and perceptual metrics---including questionnaire responses, entropy, comment length, comment ratings, and distributions of reaction types---we used nonparametric bootstrapping with resampling to estimate uncertainty around group-level means. To test for differences between treatment and control groups, we used permutation tests for Likert-scale and ordinal responses, and two-sided \textit{t}-tests for continuous measures. 

We also performed regression analyses to estimate the impact of treatment conditions and participant demographics on behavioral and perceptual measures. First, for each treatment group, we fitted a separate generalized linear model with a binomial distribution to assess the likelihood of a comment receiving a reply. Independent variables capture discussion dynamics at the time of commenting, including topic (categorical), normalized time remaining in the discussion, comment depth, number of active users (defined as users who had posted at least once), number of prior comments, and whether the AI tool was used. Second, we conducted a series of ordinary least squares regressions to examine how demographic characteristics---age ($\geq$ 45), gender (female or not), education (college degree or not), occupation (full-time employment or not), and political affiliation (Republican, Independent, Democrat)---as well as treatment group, predicted two types of outcome variables: (1) differences between pre- and post-study responses to questions on \textit{AI Related to Social Media}, and (2) user ratings of replies to own comments. Detailed information and statistical results are reported under ``Supplementary Material" in \textit{Supplementary Information}.

\subsection{AI Usage Analyses}

To better understand how participants use the AI tools, we developed a structured classification pipeline tailored to each interaction. For \textit{Chat}, \textit{Conversation Starter}, and \textit{Feedback}, we first constructed taxonomies of typical uses based on manual inspection of user interactions. We then used OpenAI's \texttt{o3-mini-2025-01-31} to classify the individual tool uses according to these taxonomies.

For the \textit{Chat} tool, we inspected submitted user-prompts and construct a taxonomy of eight broad prompt types---\textit{casual queries}, \textit{fact checking}, \textit{engagement}, \textit{political discussions}, \textit{how-to requests}, \textit{argumentation}, \textit{sentiment and context analysis}, \textit{conspiracy}---plus an \textit{other} category, each with a short description. The LLM then classified all prompts into these categories. 

A similar approach was used for the \textit{Conversation Starter}: an LLM classified which of the AI-generated conversational suggestions directly inspired a user's submitted comment. The conversation starters were categorized into ten broader themes---\textit{practical advice and suggestions}, \textit{personal experiences and anecdotes}, \textit{animal behavior and intelligence}, \textit{research and science discussions}, \textit{reflections}, \textit{debating}, \textit{comparisons}, \textit{humor}, \textit{sharing thoughts}, \textit{questions}, and \textit{other}. 

For the \textit{Feedback} tool, we identified how participants revised their comments in response to the AI feedback. We defined a taxonomy of seven revision types---\textit{structural changes}, \textit{informational updates}, \textit{argumentation}, \textit{lexical changes}, \textit{engagement enhancements}, \textit{stylistic adjustments}---plus categories for \textit{other} and \textit{no change}. The LLM classified the type of revision by comparing the user's initial draft the submitted comment. 

Finally, for the \textit{Suggestions} tool, we directly logged which of the three AI suggestion stances (agree, neutral, disagree) the participants selected. This classification framework formed the basis of the usage patterns reported in Fig.~\ref{fig:chat-conversation}a-d and~\ref{fig:feedback-suggestions}e-i. Full prompt structures and category definitions are detailed under ``Supplementary Materials" in \textit{Supplementary Information}.

\backmatter

\section{Acknowledgments}

We acknowledge the support from the Carlsberg Foundation through the COCOONS project (CF21-0432) and the National Science Foundation through Grant No. IIS-2143529. We thank the Network, Data, and Society (NERDS) group at IT University of Copenhagen, the Blablablab, and the Romero group at the School of Information, University of Michigan, for valuable feedback during internal testing.

\section{Author Contributions}

A.G.M, D.R., D.J., and L.M.A. designed the research. A.G.M. developed the platform, and collected and analyzed the data. A.G.M, D.R., D.J., and L.M.A. wrote the paper. 

\section{Competing Interests}

We declare no competing interests.

\clearpage
\appendix
\section*{Supplementary Information}
\section*{Contents}
\begin{itemize}
  \item \hyperref[sec:overview]{Supplementary Information Overview}
  \item \hyperref[sec:questionnaire-instruments]{Questionnaire Instruments}
  \item \hyperref[sec:instruments:userinstrucitons]{User Instructions}
  \item \hyperref[sec:ai-settings]{AI Prompts and Settings}
  \item \hyperref[sec:materials]{Supplementary Materials}
  \item \hyperref[sec:figures]{Supplementary Figures}
\end{itemize}

\section*{Supplementary Information Overview}
\label{sec:overview}
The Supplementary Information provides detailed documentation of the experimental workflow, technical implementation of AI tools, and our analytical approach used in the study. It includes all study materials: the informed consent form, study introduction page, pre- and post-study questionnaires, and participant instructions. We also describe the AI prompts and system settings used in each treatment condition, along with the methods used to analyze AI usage patterns, participation equality, and statistical tests. Further sections include regression model specifications for reply likelihood and demographic analyses, as well as distributions of responses to pre- and post-study questions. 

\section*{Consent Form}
\label{sec:consent-form}

Prior to participation, users were required to read and accept an informed consent form.\\

\noindent
\textbf{Eligibility}\\
Participants must be 18 years old or older.

\noindent
\textbf{Confidentiality}\\
Your data will be kept confidential and used solely for research purposes. All information will be anonymized to protect your identity. Data will be stored securely and only accessible to the research team.

\noindent
\textbf{Voluntary Participation}\\
Your participation in this study is entirely voluntary. You are free to withdraw at any time.

\noindent
\textbf{Contact Information}\\
If you have any questions or concerns about this study, please contact:
\begin{itemize}
    \item Name: Anders Giovanni M\o{}ller
    \item Email: andermol@umich.edu
\end{itemize}

\noindent
If you have any questions regarding your rights as a participant, you may contact the Institutional Review Board (IRB) of University of Michigan by email at hrppumich@umich.edu.

\noindent
\textbf{Consent Statement}\\
I have read and understood the above information, and I consent to participate in this study.

\section*{Questionnaire Instruments}
\label{sec:questionnaire-instruments}

Following the acceptance of the informed consent form, participants proceed through a structured pipeline prior to engaging in discussions. First, they read an introduction page briefly outlining the study's purpose, procedures, expected time commitment, compensation, and potential risks. Then they complete a pre-study questionnaire comprising three sections: (1) demographic information, (2) social media usage and attitudes, and (3) perceptions of AI in a social media context. After the questionnaire, participants are shown an introduction page providing guidance on their task and how the platform works. For those in the treatment groups, the page also include an explanation and demonstration of their AI tool.

Upon completion of the discussion phase, participants are directed to a post-study questionnaire. This final questionnaire includes questions evaluating their experience on the platform, their perceptions of others in the discussion, and the quality of responses received to their own comments. In the sections below, we outline the content of the pages and questionnaires.  

\subsection*{Introduction Page}
\label{sec:instruments:intropage}
Welcome to Our Social Media Engagement Study!\\

\noindent
\textbf{Purpose of the Study}\\
We are conducting a study to explore how participants engage with different types of social media posts of various topics.

\noindent
\textbf{What You'll Do}\\
If you choose to participate in this study, this is what you will do:

\begin{itemize}
    \item You will be placed in a room with [N] other participants and exposed to a social media platform.
    \item During the experiment, you will see a live social media conversation with a post and a few comments. You and the other participants will have 10 minutes to interact and engage with the content by liking and commenting.
    \item This process will be repeated for three different posts, one at a time.
    \item Before and after the experiment, you will be asked to answer some questions.
\end{itemize}

\noindent
\textbf{ChatGPT/AI Assistance}\\
Throughout the experiment, you are NOT ALLOWED to use any EXTERNAL ChatGPT/AI-assisted functionality.

\noindent
\textbf{Your Participation and Compensation}\\
Your participation is completely voluntary, and you can choose to stop at any time without any consequences. You will receive your payment upon completing the final questionnaire.

\noindent
\textbf{Benefits and Risks}\\
There are no anticipated risks beyond those encountered in everyday social media use. You might encounter posts and content that include misinformation or opinions that differ from your own. You might also experience exposure to negative or toxic content.

\noindent
\textbf{Time Commitment}\\
The study will take approximately 30-40 minutes to complete.

\subsection*{Questionnaires}
\phantomsection
\label{sec:instruments:questionnairequestions}

\subsubsection*{Pre-study Questionnaire}

\noindent
\textbf{Demographic Questions}
\begin{itemize}
  \item \textit{What is your age group?} Options: 18-24, 25-34, 35-44, 45-54, 55-64, 65+, Prefer not to answer. 
  \item \textit{How do you identify your gender?} Options: male, female, non-binary, prefer not to answer.
  \item \textit{What is the highest level of education you have completed?} Options: did not graduate from high school, high school graduate, some college but no degree, 2-year college degree, 4-year college degree, postgraduate degree (MA, MBA, JD, PhD, etc), prefer not to answer.
  \item \textit{What is your current occupation or employment status?} Options: student, employed full-time, employed part-time, self-employed, unemployed, retired, other (please specify), prefer not to answer.
  \item \textit{Which of the following best describes your political party affiliation or the political group you most closely align with? Please choose the option that best reflects your stance, even if you don't fully agree with everything that party stands for.} Options: strong democrat, moderate democrat, independent, moderate republican, strong republican, other (libetarian, green party, etc.), prefer not to answer. 
\end{itemize}

\noindent
\textbf{Social Media Questions}
\begin{itemize}
    \item \textit{Which social media platforms do you use most frequently? (Select all that apply)}. Options: Facebook, Twitter (X), Instagram, LinkedIn, Tiktok, Reddit, Bluesky, Mastodon, Snapchat, other (please specify), prefer not to answer.
    \item \textit{I seek information about health and wellbeing practices on online social media platforms.} Options: 5-point Likert scale.
    \item \textit{I stay informed about current events and issues in US politics.} Options: 5-point Likert scale.
    \item \textit{I often participate in online discussions on social media.} Options: 5-point Likert scale.
    \item \textit{It is easy for me to engage in online conversations.} Options: 5-point Likert scale.
    \item \textit{I find discussions on social media informative and high-quality.} Options: 5-point Likert scale.
    \item \textit{I trust the information shared by other users on social media.} Options: 5-point Likert scale
    \item \textit{I recall changing my opinion based on interactions on social media.} Options 5-point Likert scale.
    \item \textit{I tend to agree with the opinions I see on social media.} Options: 5-point Likert scale.
    \item \textit{Do you face any barriers when posting content on social media? (Select all that apply).} Options: None, lack of ideas, time constraint, fear of negative feedback, technical difficulties, privacy concerns, other (please specify), prefer not to answer.
\end{itemize}

\noindent
\textbf{AI Related to Social Media Questions}
\begin{itemize}
    \item \textit{I feel comfortable with AI being used on social media platforms.} Options: 5-point Likert scale.
    \item \textit{AI suggestions can make it more likely for me to participate in online discussions.} Options: 5-point Likert scale.
    \item \textit{AI can make online discussions more positive and less toxic.} Options: 5-point Likert scale.
    \item \textit{AI can make discussions less polarizing.} Options: 5-point Likert scale.
    \item \textit{AI can help reduce misinformation on social media.} Options: 5-point Likert scale.
    \item \textit{AI-generated content is accurate and reliable.} Options: 5-point Likert scale.
    \item \textit{AI should be regulated to prevent misuse and ensure ethical use.} Options: 5-point Likert scale.
\end{itemize}

\subsubsection*{Post-study Questionnaire}

\noindent
\textbf{Platform Experience}
\begin{itemize}
  \item \textit{Rate your overall experience with the platform}. Options: Very poor, poor, fair, good, excellent, Prefer not to answer. 
  \item \textit{I participated in the discussions more than I usually do on social media.} Options: 5-point Likert-scale.
  \item \textit{It was easy for me to engage in the conversations.} Options: 5-point Likert-scale.
  \item \textit{I found the comments from other participants to be informative and high-quality.} Options: 5-point Likert-scale.
  \item \textit{I trust the information provided by other participants.} Options: 5-point Likert-scale.
  \item \textit{I tended to agree with the other participants.} Options: 5-point Likert-scale.
  \item \textit{I feel like I had enough time for each conversation.} Options: 5-point Likert-scale.
  \item \textit{Did you face any barriers when posting content?} Options: None, lack of ideas, time constraints, fear of negative feedback, technical difficulties, privacy concerns, other (please specify).
  \item \textit{General feedback about the platform.} Open-ended input.
\end{itemize}

\noindent
\textbf{AI Evaluation (only for treatment participants)}
\begin{itemize}
  \item \textit{The AI made it easier for me to participate in the discussion, compared to my usual experience on social media.} Options: 5-point Likert-scale.
  \item \textit{The AI felt natural in the context of the discussions.} Options: 5-point Likert-scale.
  \item \textit{The content I created using the AI was high-quality.} Options: 5-point Likert-scale.
  \item \textit{I can imagine situations in which I would use this AI if it was available on social media.} Options: 5-point Likert-scale.
  \item \textit{Any other feedback you would like to share about the AI.} Open-ended input.
\end{itemize}

\noindent
\textbf{AI Related to Social Media Questions (similar to the initial questionnaire items)}
\begin{itemize}
    \item \textit{I feel comfortable with AI being used on social media platforms.} Options: 5-point Likert scale.
    \item \textit{AI suggestions can make it more likely for me to participate in online discussions.} Options: 5-point Likert scale.
    \item \textit{AI can make online discussions more positive and less toxic.} Options: 5-point Likert scale.
    \item \textit{AI can make discussions less polarizing.} Options: 5-point Likert scale.
    \item \textit{AI can help reduce misinformation on social media.} Options: 5-point Likert scale.
    \item \textit{AI-generated content is accurate and reliable.} Options: 5-point Likert scale.
    \item \textit{AI should be regulated to prevent misuse and ensure ethical use.} Options: 5-point Likert scale.
\end{itemize}

\noindent
\textbf{Ratings of users and comments}
\begin{itemize}
    \item \textbf{Comments}
    \begin{itemize}
        \item \textit{Rate the comment's value to the discussion. Please rate the following comments from 1 (low) to 5 (high), with higher meaning better.} We sample up to 10 answers to a users own comments. 
    \end{itemize}
    \item \textbf{Users}: We ask participants to rate the other users they engaged with. For numerical options are 1 representing low and 5 high.
    \begin{itemize}
        \item \textit{Rate the user's positivity.} Options: 1 to 5 scale.
        \item \textit{Rate the user's engagement.} Options: 1 to 5 scale.
        \item \textit{Rate the user's politeness.} Options: 1 to 5 scale.
        \item \textit{Rate the user's political affiliation.} Options: democrat, republican, other, don't know.
        \item \textit{Do you feel like you share the same values with the user?} Options: yes, no, don't know.
        \item \textit{Do you think the user is a bot?} Options: yes, no, don't know.
        \item \textit{Do you think the user used AI assistance?} Options: yes, no, don't know.
    \end{itemize}
\end{itemize}

\section*{User Instructions}
\label{sec:instruments:userinstrucitons}

\subsection*{Platform Introduction}

This section provides an overview of the platform's interface. Each conversation will contain content for you to interact with, and other users will also be actively engaging. You will see an initial post accompanied by comments. You can interact by clicking the ``Comment" button to add a comment or the ``React" button to express a reaction.

\noindent
\textbf{Instructions}
\begin{itemize}
    \item Once the timer here ends, you will be presented with a social media-like platform, as shown on the right.
    \item First, carefully read the content shown when you enter the conversation.
    \item Your task is to engage with the content by expressing your opinions, commenting on posts, and reacting to them.
    \item You will have 10 minutes to interact with each conversation.
    \item There are a total of 3 conversations to participate in.
\end{itemize}

\noindent
Participants in any of the treatment groups will see the following AI-dependent information:

\subsection*{Chat}

\noindent
\textbf{AI Feature}\\
You have been selected to try our AI Assistant tool.

On the right side of your screen, you'll find a chat window where you can interact with the assistant. Feel free to ask about the content, request suggestions for what to write, pose clarifying questions, and more. The assistant is here to support you as you engage with the material.

\subsection*{Suggestions}

\noindent
\textbf{AI Feature}\\
You have been selected to try our AI Suggestions tool.

The AI Suggestions feature provides you with three "ready to use" comments for the post you want to respond to. When you activate this feature, you'll receive three different comment options:
\begin{itemize}
    \item A comment that agrees with the post's stance.
    \item A neutral comment.
    \item A comment that disagrees with the post's stance.
\end{itemize}

You can choose to use one of these suggestions as-is or use them as a starting point for your own comment. This feature is designed to help you engage with different perspectives and can be particularly useful when you're unsure how to approach a topic or want to explore various viewpoints.

\subsection*{Feedback}

\noindent
\textbf{AI Feature}\\
You have been selected to try our AI Feedback tool.

With the AI Feedback feature, you can get assistance to enhance your comments when responding to posts or other comments. After you've written your response, you can ask the AI to provide reflections and suggestions on various aspects of your comment. This feedback can help you improve the quality, clarity, and effectiveness of your contributions to the discussion. Feel free to use these insights to refine your thoughts and engage more meaningfully with the content and other users.

\subsection*{Conversation Starter}

\noindent
\textbf{AI Feature}\\
You have been selected to try our AI Conversation Starter tool.

Next to the comment button, you'll notice a light purple button - this is our AI Conversation Starter. When you click on it, our AI will provide thoughts and ideas on how to start a conversation about the particular post or comment you're viewing. This feature is designed to help you engage more easily, especially when you want to participate but aren't sure how to begin. It's a great tool for overcoming writer's block or simply getting inspiration for your responses. Feel free to use these AI-generated suggestions as a starting point for your own unique contributions to the discussion.

\section*{AI Prompts and Settings}
\label{sec:ai-settings}

To enable our AI tools, we construct custom prompts tailored to the intended function of each tool.

\subsection*{Model Configuration and API Call Settings}

We used the OpenAI API via the official JavaScript client (\texttt{openai} npm package) to interact with the AI across treatment conditions with an individual prompt for each of the tools. All treatments use the same parameters to ensure consistency. Below is an overview of the configuration.

\begin{itemize}
    \item Model: \texttt{gpt-4o}
    \item Temperature: $1.0$
    \item Max\_tokens: $1000$
    \item seed: \texttt{Math.floor(Math.random() * 1000000)}
\end{itemize}

\noindent
Example of API call for the \textit{Chat} feature:

\begin{JSCodeBox}
const openai = new OpenAI({
    apiKey: OPENAI_API_KEY
});

const response = await openai.chat.completions.create({
    model: "gpt-4o",
    messages: [
        { role: "system", content: systemPrompt },
       ...recentMessages, // If any prior messages
        { role: "user", content: userInput }
    ],
    max_tokens: 1000,
    temperature: 1.0,
    seed: Math.floor(Math.random() * 1000000),
});
\end{JSCodeBox}

We set the model temperature parameter to $1.0$ to increase creativity and variability in the AI-generated content. We use random seeds for each API call, ensuring varied outputs. 

\subsection*{Chat}

\begin{tcolorbox}[title=Chat Prompt,fonttitle=\bfseries]
\ttfamily
You are a highly engaging and informative AI designed to help users increase their interest in various topics, improve their understanding, increase their trust in reliable information, and encourage them to interact with content by liking and commenting.

The format of the posts and their nested comments is a readable text structure, where each post starts with its author and content, followed by likes, and then displays all comments with proper indentation (using spaces) to show the comment hierarchy, with each comment showing its author, content, and likes. 

Content:
\{content\}

Use the above content, the user's behavior, and the demographic context to answer questions related to the content and the topic in an engaging and informative manner. Keep the answers short and effective in communication to a broad audience in a few sentences.

Instructions:

- Write in a natural, conversational tone, as if you're a real person engaging on social media.

- Avoid overly formal language or technical jargon.

- The responses don't have to be overly respectfully, they can be direct and informal.

- Reflect typical social media interaction styles, including informal punctuation if suitable.

- Be aware of the comments and likes from the user, it is indicated by "You".

\end{tcolorbox}

After the defined system message above, we add any potential prior messages of the AI-human conversation and the specific user prompt. 

\subsection*{Suggestions}

\begin{tcolorbox}[title=Suggestion Prompt,fonttitle=\bfseries]
\ttfamily
You are a highly engaging and informative AI designed to help users increase their interest in various topics, improve their understanding, increase their trust in reliable information, and encourage them to interact with content by liking and commenting.

Given the following post, comments, replies, and demographic information, generate 3 diverse potential replies to a specific part of the content. Be aware of the content, the user's behavior (likes, comments). Use this information to generate replies that are engaging and informative and maintain the style of the content and social media in general.

The format of the posts and their nested comments is a readable text structure, where each post starts with its author and content, followed by likes, and then displays all comments with proper indentation (using spaces) to show the comment hierarchy, with each comment showing its author, content, and likes.

Content:
\{content\}

Specific Part:
\{specificPart\}

Generate 3 diverse potential replies in JSON format based on the specific part of the content:
1. A reply agreeing with the content. You don't have to say "I agree" or anything like that, just reply in a way that agrees with the content.
2. A balanced or neutral reply. You don't have to say "I think" or anything like that, just reply in a way that is neutral or balanced.
3. A reply disagreeing or challenging the content. You don't have to say "I disagree", just reply in a way that disagrees with the content.

Instructions:

- Write in a natural, conversational tone, as if you're a real person engaging on social media.

- Avoid overly formal language or technical jargon.

- The responses don't have to be overly respectfully, they can be direct and informal.

- Reflect typical social media interaction styles, including informal punctuation if suitable.

- Be aware of the comments and likes from the user, it is indicated by "You".
\end{tcolorbox}

In the context above, \texttt{specificPart} refers to the comment the users is generating replies for. 

\subsection*{Feedback}

\begin{tcolorbox}[title=Feedback Prompt,fonttitle=\bfseries]
\ttfamily
You are an AI assistant designed to help users improve their social media comments by providing friendly, constructive feedback. Your goal is to offer insights that can help the user refine their comment, making it more engaging, clear, and appropriate for the situation.

Context:
A user is about to post a comment on a social media platform like Reddit. They want to ensure their comment is effective and adds value to the conversation.

Original Post:
\{originalPost\}

User's Comment:
\{userComment\}

Your Task:
Provide helpful feedback on the user's comment, considering the context of the original post. Your feedback should feel like it's coming from a real person and should be dynamic, varying in length and content based on the specific situation. Include any insights that you think would be most helpful to the user in improving their comment. This might involve:

- Offering suggestions for improvement or clarification.

- Providing examples or alternative phrasings to illustrate your suggestions.

- Asking questions that encourage deeper thinking or alternative perspectives.

- Relating personal experiences or reflections that are relevant.

- Pointing out any potential issues with tone, clarity, or appropriateness.

- Highlighting inconsistencies or contradictions and suggesting corrections.

- Suggesting ways to enhance engagement or make the comment more impactful.

- Noting any potential misunderstandings and offering clarifications.

- Encouraging the inclusion of evidence or sources if appropriate.

- Advising on language and terminology to suit the audience.

- Considering the audience's perspective and suggesting adjustments.

- Promoting empathy and understanding to foster respectful dialogue.

- Highlighting opportunities for humor or creativity, if suitable.

- Any other observations that could help the user enhance their comment.

Response Format:
Present your feedback in a JSON object. Each key should be a brief descriptive title of your feedback point, and the value should be your detailed feedback message. The number and types of feedback points should vary depending on what you think is most helpful for the user in this context.

Instructions:

- Write in a friendly, conversational tone, as if you're chatting with a friend.

- Use natural, everyday language; avoid sounding robotic or overly formal.

- Be genuine and personable in your feedback.

- Tailor your feedback specifically to the user's comment and the context.

- The number of feedback points can vary; it's not necessary to include all possible types.
\end{tcolorbox}

\subsection*{Conversation Starter}

\begin{tcolorbox}[title=Conversation Starter Prompt,fonttitle=\bfseries]
\ttfamily

You are an AI assistant designed to help users engage in conversations on social media platforms like Reddit. Your task is to suggest different approaches for starting engaging conversations. Think of it as coaching a friend on different ways they could approach the conversation, without giving them specific words to use.

The content consists of posts and their nested comments presented in a readable text structure. Each post starts with its author and content, followed by likes. Comments are indented to show the comment hierarchy, with each comment showing its author, content, and likes.

Content:
\{content\}

Comment to reply to:
\{replyToComment\}

Response Format:
Based on the content and the comment to reply to, generate several ideas for conversation starters that the user can use to get started engaging in the discussion. Each suggestion should have a descriptive title, and the conversation starter should be natural, relevant, and encourage further conversation without sounding robotic or AI-like.

Output your response as a JSON object.

Instructions:

- Write in a natural, conversational tone, as if you're a real person engaging on social media.

- Avoid overly formal language or technical jargon.

- Be direct and informal; it's okay to use colloquial expressions, slang, or emojis if appropriate.

- Reflect typical social media interaction styles, including informal punctuation if suitable.

- Ensure that the ideas are diverse and cover different angles or perspectives on the topic.

- Do not mention any objectives or categories in your responses; just provide ideas for conversation starters.
\end{tcolorbox}
\

\section*{Supplementary Material}
\label{sec:materials}

\subsection*{Statistical Tests and Bootstrapping}

We employ bootstrapping with $10,000$ iterations, using resampling with replacement to estimate uncertainty of our observed metrics. We obtain percentile-based confidence intervals by extracting 2.5th and 97.5th percentiles from the bootstrapped distributions. We use a fixed seed for reproducibility. For statistical comparisons between control and treatment groups, we use two methods depending on data type: permutation tests for Likert-scale and ordinal responses (excluding treatment effects using pre- and post-study answers), and two-sided \textit{t}-tests for continuous or normally distributed metrics. Across all tests, we use the following significance markers: \textdagger\textit{p} $<$ 0.10; *\textit{p} $<$ 0.05; **\textit{p} $<$ 0.01; ***\textit{p} $<$ 0.001.

\subsection*{AI Usage}

To better understand how participants used the AI tools, we developed structured classification pipelines specific to the design of each intervention. For tools involving free-form interaction (\textit{Chat}, \textit{Conversation Starter}, \textit{Feedback}), we first define a taxonomy of use cases or content change through manual inspection. We then use OpenAI's API and model \texttt{o3-mini-2025-01-31} to classify individual tool usage according to these taxonomies. 

\subsubsection*{Chat}

For the \textit{Chat} tool, we define the following taxonomy through manual review of the user prompts to the AI assistant. 

\begin{itemize}
    \item \textbf{Casual queries}: Users engage in open-ended, informal, and sometimes playful discussions without a clear goal beyond conversation.
    \item \textbf{Fact checking}: Users request research-backed answers, seek validation for claims, and debunk misinformation.
    \item \textbf{Engagement}: Users ask for help crafting responses, engaging in social interactions, and making witty or humorous replies.
    \item \textbf{Political discussions}: Users seek analysis of policies, societal structures, ethical dilemmas, and political theories.
    \item \textbf{How-to requests}: Users ask for step-by-step guides, explanations, or structured insights.
    \item \textbf{Argumentation}: Users seek structured arguments, counterpoints, and persuasive responses for discussions and debates.
    \item \textbf{Sentiment and context analysis}: Users seek summaries, context, or sentiment analysis of discussions.
    \item \textbf{Conspiracy}: Users engage in speculation, conspiracy theories, and alternative explanations.
    \item \textbf{Other}: None of the above applies.
\end{itemize}

Using this taxonomy, we apply the following prompt to make the classification:

\begin{tcolorbox}[title=Chat usage prompt,fonttitle=\bfseries]
\ttfamily

You are a helpful assistant that classifies chat messages into one of the following message categories. Each message category contains a brief description of the category.

Message Categories:

\{categories\}

If none of the message categories apply, classify the message as "other".

Message: \{message\}

Category:
\end{tcolorbox}











\subsubsection*{Conversation Starter}

To understand how the AI-generated conversation starters influences user behavior, we first extract all suggestions shown to participants, which contained a ``starter type'' and corresponding description or suggestion. We group these types broader categories. Each submitted comment where the conversation starter was used is categorized into these categories.

\begin{itemize}
    \item \textbf{Practical advice and suggestions}: Providing actionable advice, tips, practical ideas, and health-related suggestions.
    \item \textbf{Personal experiences and anecdotes}: Encouraging users to share personal stories, experiences, or reflections.
    \item \textbf{Animal behavior and intelligence}: Exploring and discussing animal-related behaviors, training, intelligence, and cross-species comparisons.
    \item \textbf{Research and science discussions}: Focused on empirical research findings, scientific insights, validity, and methodological exploration.
    \item \textbf{Reflections}: Inviting philosophical reflection, hypothetical scenarios, and thoughtful speculation on past, present, or future topics.
    \item \textbf{Debating}: Stimulating critical thinking by challenging viewpoints, initiating friendly debates, or questioning assumptions.
    \item \textbf{Comparisons}: Encouraging analytical thinking through direct comparisons, contrasts, and relational insights.
    \item \textbf{Humor}: Using humor and playful interactions to foster engagement and create positive interactions. 
    \item \textbf{Sharing thoughts}: Suggestions related to sharing and expressing thoughts.
    \item \textbf{Questions}: Engaging users by prompting questions, exploring topics, or encouraging deeper discussions.
    \item \textbf{Other}: None of the above applies.
\end{itemize}

Using these categories and descriptions, we apply the following prompt for the classification.

\begin{tcolorbox}[title=Conversation Starter usage prompt,fonttitle=\bfseries]
\ttfamily

You are a helpful assistant that identifies which conversation starters influenced a social media comment.

Conversation starters:
\{categories\}

For the comment below, list any conversation starters that appear to have directly influenced or inspired the comment. If the comment doesn't seem influenced by any of the listed conversation starters, classify it as "other".

Comment: \{comment\}

Conversation starters that directly influenced this comment:
\end{tcolorbox}

\subsubsection*{Feedback}

To understand how participants used the Feedback tool, we compared pairs of draft and submitted comments. We define a taxonomy capturing the types of textual changes after receiving AI-generated feedback. 

\begin{itemize}
    \item \textbf{Structural changes}: Modifications in sentence structure, paragraph flow, or content length.
    \item \textbf{Informational updates}: Corrections and additions that improve factual accuracy and completeness.
    \item \textbf{Argumentation}: Improvements in reasoning, logic, and argument strength.
    \item \textbf{Lexical changes}: Word-level modifications for clarity, precision, and readability.
    \item \textbf{Engagement enhancements}: Refinements that improve engagement, rhetorical impact, and readability. 
    \item \textbf{Stylistic adjustments}: Changes in tone, formality, or politeness of the text.
    \item \textbf{Sentiment adjustments}: Changes that affect the emotional tone and expressiveness of the text.
    \item \textbf{Other}: None of the above applies.
    \item \textbf{No changes}: The user did not change they draft comment after receiving feedback.
\end{itemize}

The categories and descriptions are used in the following classification prompt, omitting pairs with no changes in the text after receiving feedback:

\begin{tcolorbox}[title=Feedback usage prompt,fonttitle=\bfseries]
\ttfamily

You are a helpful assistant that classifies changes between two versions of a text into one of the following change categories. Each change category contains a brief description of the category.

Change Categories:

\{categories\}

If none of the categories apply, classify the change as "other".

Original Text: \{original\_text\}

Revised Text: \{revised\_text\}

Change Pattern:
\end{tcolorbox}

Additionally, we measure the similarity between comment drafts before and after receiving AI-based feedback using Jaccard similarity, computed as the intersection over union of the words in the two texts. Higher values indicate greater lexical overlap and lower values reflect fewer shared words.

\subsubsection*{Suggestions}

The \textit{Suggestions} tool offered participants three AI-generated reply-options, one agreeing, one neutral, and one disagreeing with the comment they were responding to. Users could select one of these responses or ignore the suggestions. For this analysis, we directly inspect which suggestion was selected.

In addition, we test whether the distribution of user selections (agree, neutral, disagree) differs by topic. Specifically, we conduct Chi-square tests of independence to analyze whether selections are associated with topic. We perform two pairwise comparisons: \textit{cats} vs. \textit{oats} and \textit{cats} vs. \textit{politics}. These tests are based on the following contingency tables: 

\begin{table}[htp]
\centering
\begin{tabular}{lrl}
\toprule
\textbf{Selection} & \textbf{\textit{cats} (n=X)} & \textbf{\textit{oats} (n=Y)} \\
\midrule
Agree & 168 & 216 \\
Neutral & 130 & 117 \\
Disagree & 90 & 90 \\
\bottomrule
\end{tabular}
\caption{Contingency table of users selections for \textit{cats} and \textit{oats}.}
\label{tab:suggestions-cats-oats}
\end{table}

\noindent
Chi-square statistic: $\chi^2(\text{df=2})=5.18, p=0.075$.

\begin{table}[htp]
\centering
\begin{tabular}{lrl}
\toprule
\textbf{Selection} & \textbf{\textit{cats} (n=X)} & \textbf{\textit{politics} (n=Y)} \\
\midrule
Agree & 168 & 198 \\
Neutral & 130 & 112 \\
Disagree & 90 & 76 \\
\bottomrule
\end{tabular}
\caption{Contingency table of users selections for \textit{cats} and \textit{politics}.}
\label{tab:suggestions-cats-oats}
\end{table}

\noindent
Chi-square statistic: $\chi^2(\text{df=2})=4.97, p=0.083$.

\subsection*{Post-hoc Regression Analyses on Treatment Impact of Reply Likelihood}

To better understand how the different AI-tool interventions influence the likelihood that a comment receives a reply, we conduct a set of post-hoc regression analyses across the four treatment conditions. For each treatment group, we fit a separate generalized linear model (GLM) with a binomial distribution. The dependent variables is a binary indicator (has\_answers) denoting whether a comment received at least one direct reply. 

Each regression model include consistent independent variables chosen to reflect structural factors of the discussion dynamics at the time of commenting. 

\begin{itemize}
    \item \textbf{Topic} (\texttt{bool}): Categorical variable indicating the discussion topic, with the \textit{Cats} topic omitted as reference to avoid multicollinearity.
    \item \textbf{Time left} (\texttt{float}): Continuous and normalized variable indicating how much time remained in the 10-minute discussion round when the comment was posted.
    \item \textbf{Depth} (\texttt{int}): The depth of the comment within the tread.
    \item \textbf{Active users} (\texttt{int}): The number of unique participants who have posted at least one comment at the time of the comment.
    \item \textbf{Treatment used} (\texttt{bool}): Binary indicator of whether the AI tool was used before the comment was submitted. Here the variable is dependent on the design of the tool. We set the following logic for the treatments; \textit{Chat}: If the user has interacted with the AI-chat within two minutes prior to submitted a comment. \textit{Conversation Starter}: Did the user use the Conversation Starter on the parent comment. \textit{Feedback}: Did the user review feedback before submitting the comment. \textit{Suggestions:} Did the user receive and select a Suggestion before submitting a comment. Here we also test if the specific Suggestion type, i.e. agree, neutral, disagree, would have an impact.
    \item \textbf{Number of prior comments} (\texttt{int}): Total number of comments submitted in the round. 
\end{itemize}

Below we report the results for each treatment group.

\begin{table}[htp]
\centering
\begin{tabular}{lrrrrl}
\toprule
Variable & Coef & Std. Err. & z & p-value & 95\% CI \\
\midrule
Intercept & -0.798 & 0.445 & -1.790 & 0.073 & [-1.670, 0.075] \\
C(topic)[T.oats] & 0.347 & 0.144 & 2.410 & 0.016 & [0.065, 0.629] \\
C(topic)[T.politics] & 0.283 & 0.150 & 1.890 & 0.058 & [-0.010, 0.576] \\
time\_left & 0.933 & 0.477 & 1.960 & 0.051 & [-0.002, 1.868] \\
depth & 0.061 & 0.060 & 1.030 & 0.305 & [-0.056, 0.178] \\
active\_users & 0.107 & 0.060 & 1.770 & 0.076 & [-0.011, 0.226] \\
treatment\_used & 0.053 & 0.121 & 0.440 & 0.660 & [-0.184, 0.290] \\
n\_comments\_before & -0.083 & 0.025 & -3.310 & 0.001 & [-0.133, -0.034] \\
\bottomrule
\end{tabular}
\caption{Logistic regression results for the \textbf{Chat} treatment condition.}
\end{table}

\begin{table}[htp]
\centering
\begin{tabular}{lrrrrl}
\toprule
Variable & Coef & Std. Err. & z & p-value & 95\% CI \\
\midrule
Intercept & -1.946 & 0.349 & -5.581 & $<$0.001 & [-2.630, -1.263] \\
C(topic)[T.oats] & 0.236 & 0.140 & 1.689 & 0.091 & [-0.038, 0.510] \\
C(topic)[T.politics] & 0.384 & 0.140 & 2.737 & 0.006 & [0.109, 0.659] \\
time\_left & 1.927 & 0.363 & 5.306 & $<$0.001 & [1.215, 2.639] \\
depth & 0.072 & 0.059 & 1.222 & 0.222 & [-0.044, 0.188] \\
active\_users & 0.089 & 0.059 & 1.512 & 0.131 & [-0.026, 0.203] \\
treatment\_used & 0.300 & 0.130 & 2.304 & 0.021 & [0.045, 0.555] \\
n\_comments\_before & -0.007 & 0.018 & -0.370 & 0.711 & [-0.043, 0.029] \\
\bottomrule
\end{tabular}
\caption{Logistic regression results for the \textbf{Conversation Starter} treatment condition.}
\end{table}

\begin{table}[htp]
\centering
\begin{tabular}{lrrrrl}
\toprule
Variable & Coef & Std. Err. & z & p-value & 95\% CI \\
\midrule
Intercept & -3.098 & 0.444 & -6.985 & $<$0.001 & [-3.967, -2.229] \\
C(topic)[T.oats] & 0.378 & 0.165 & 2.286 & 0.022 & [0.054, 0.702] \\
C(topic)[T.politics] & 0.395 & 0.173 & 2.281 & 0.023 & [0.056, 0.733] \\
time\_left & 3.529 & 0.471 & 7.495 & $<$0.001 & [2.606, 4.452] \\
depth & 0.059 & 0.089 & 0.662 & 0.508 & [-0.116, 0.234] \\
active\_users & -0.009 & 0.070 & -0.124 & 0.901 & [-0.146, 0.128] \\
treatment\_used & 0.243 & 0.153 & 1.588 & 0.112 & [-0.057, 0.542] \\
n\_comments\_before & 0.050 & 0.023 & 2.162 & 0.031 & [0.005, 0.096] \\
\bottomrule
\end{tabular}
\caption{Logistic regression results for the \textbf{Feedback} treatment condition.}
\end{table}

\begin{table}[htp]
\centering
\begin{tabular}{lrrrrl}
\toprule
Variable & Coef & Std. Err. & z & p-value & 95\% CI \\
\midrule
Intercept & -2.387 & 0.333 & -7.176 & $<$0.001 & [-3.039, -1.735] \\
C(topic)[T.oats] & 0.227 & 0.120 & 1.891 & 0.059 & [-0.008, 0.463] \\
C(topic)[T.politics] & 0.144 & 0.125 & 1.157 & 0.247 & [-0.100, 0.388] \\
time\_left & 2.835 & 0.318 & 8.921 & $<$0.001 & [2.212, 3.458] \\
depth & 0.154 & 0.030 & 5.208 & $<$0.001 & [0.096, 0.212] \\
active\_users & 0.117 & 0.048 & 2.437 & 0.015 & [0.023, 0.212] \\
treatment\_used & 0.045 & 0.105 & 0.428 & 0.669 & [-0.162, 0.252] \\
n\_comments\_before & -0.004 & 0.008 & -0.501 & 0.616 & [-0.019, 0.011] \\
\bottomrule
\end{tabular}
\caption{Logistic regression results for the \textbf{Suggestions} treatment condition.}
\end{table}

\FloatBarrier

\subsection*{Demographic Regression Analysis on questionnaire-based Variables}

To examine whether demographic background influenced how participants evaluated elements of the study or perceived the effects of AI interventions, we conducted a series of ordinary least squares (OLS) regressions. The dependent variables (DVs) are within two distinct categories:

\begin{enumerate}
    \item \textbf{Comment Ratings}: Average ratings of answers received to own comments.
    \item \textbf{Change in questionnaire answers ($\Delta$)}: Difference between pre- and post-study questionnaire responses for the questions on \textit{AI Related to Social Media} outlined in~\nameref{sec:instruments:questionnairequestions}.
\end{enumerate}

Each regression includes the same set of demographic-based independent variables (IVs). We process the demographic information into coarser groups. 

\begin{itemize}
    \item \textbf{Age} (\texttt{binary}): Older than 44 or not. 
    \item \textbf{Gender} (\texttt{binary}): Female or not female.
    \item \textbf{Education} (\texttt{binary}): At least a 4-year college degree or not. 
    \item \textbf{Occupation} (\texttt{binary}): Has full-time job or not. 
    \item \textbf{Political Affiliation} (\texttt{categorical}): Indicator variables for Republican and Independent with Democrat as the reference.
    \item \textbf{Treatment Group} (\texttt{binary}): One-hot encoded binary variables, one for each group.
\end{itemize}

Below, we report full regression results for each outcome variable. As noted in the main text, the regressions show limited and inconsistent effects of demographic and treatment variables on participant perceptions of AI and the content.

\begin{table}[htp]
\centering
\begin{tabular}{lrrrrr}
\toprule
Variable & Coef. & Std. Err. & t & p-value & 95\% CI \\
\midrule
Intercept & 3.613 & 0.103 & 35.027 & $<$0.001 & [3.410, 3.816] \\
treatment\_chat[T.True] & -0.179 & 0.102 & -1.758 & 0.079 & [-0.380, 0.021] \\
treatment\_conversation[T.True] & -0.293 & 0.102 & -2.874 & 0.004 & [-0.492, -0.093] \\
treatment\_feedback[T.True] & -0.089 & 0.102 & -0.867 & 0.386 & [-0.290, 0.112] \\
treatment\_suggestions[T.True] & 0.065 & 0.102 & 0.634 & 0.526 & [-0.136, 0.265] \\
age\_old[T.True] & 0.010 & 0.065 & 0.146 & 0.884 & [-0.118, 0.137] \\
gender\_not\_female[T.True] & -0.117 & 0.065 & -1.795 & 0.073 & [-0.245, 0.011] \\
education\_not\_educated[T.True] & -0.016 & 0.068 & -0.237 & 0.813 & [-0.149, 0.117] \\
occupation\_full\_time[T.True] & -0.015 & 0.068 & -0.216 & 0.829 & [-0.148, 0.119] \\
party\_Independent[T.True] & 0.061 & 0.079 & 0.764 & 0.445 & [-0.095, 0.216] \\
party\_Republican[T.True] & 0.250 & 0.078 & 3.202 & 0.001 & [0.096, 0.403] \\
\bottomrule
\end{tabular}
\caption{\emph{Average Comment Rating}.}
\end{table}

\begin{table}[htp]
\centering
\begin{tabular}{lrrrrr}
\toprule
Variable & Coef. & Std. Err. & t & p-value & 95\% CI \\
\midrule
Intercept & 0.038 & 0.117 & 0.324 & 0.746 & [-0.191, 0.267] \\
treatment\_chat[T.True] & 0.409 & 0.116 & 3.522 & $<$0.001 & [0.181, 0.636] \\
treatment\_conversation[T.True] & 0.039 & 0.116 & 0.333 & 0.739 & [-0.190, 0.267] \\
treatment\_feedback[T.True] & 0.069 & 0.116 & 0.590 & 0.555 & [-0.160, 0.298] \\
treatment\_suggestions[T.True] & 0.252 & 0.117 & 2.160 & 0.031 & [0.023, 0.482] \\
age\_old[T.True] & 0.216 & 0.075 & 2.891 & 0.004 & [0.069, 0.363] \\
gender\_not\_female[T.True] & -0.103 & 0.074 & -1.384 & 0.167 & [-0.249, 0.043] \\
education\_not\_educated[T.True] & 0.062 & 0.077 & 0.808 & 0.419 & [-0.089, 0.213] \\
occupation\_full\_time[T.True] & 0.117 & 0.077 & 1.509 & 0.132 & [-0.035, 0.269] \\
party\_Independent[T.True] & 0.003 & 0.090 & 0.034 & 0.973 & [-0.174, 0.180] \\
party\_Republican[T.True] & -0.188 & 0.089 & -2.108 & 0.035 & [-0.364, -0.013] \\
\bottomrule
\end{tabular}
\caption{Change in answers between pre- and post-study questionnaire on the question\emph{``I feel comfortable with AI being used on social media platforms.''}}
\end{table}

\begin{table}[htp]
\centering
\begin{tabular}{lrrrrr}
\toprule
Variable & Coef. & Std. Err. & t & p-value & 95\% CI \\
\midrule
Intercept & 0.262 & 0.139 & 1.886 & 0.060 & [-0.011, 0.536] \\
treatment\_chat[T.True] & 0.375 & 0.138 & 2.711 & 0.007 & [0.103, 0.647] \\
treatment\_conversation[T.True] & 0.021 & 0.139 & 0.151 & 0.880 & [-0.251, 0.293] \\
treatment\_feedback[T.True] & -0.154 & 0.139 & -1.109 & 0.268 & [-0.427, 0.119] \\
treatment\_suggestions[T.True] & 0.263 & 0.140 & 1.881 & 0.060 & [-0.012, 0.538] \\
age\_old[T.True] & -0.072 & 0.089 & -0.811 & 0.418 & [-0.248, 0.103] \\
gender\_not\_female[T.True] & -0.105 & 0.089 & -1.187 & 0.236 & [-0.279, 0.069] \\
education\_not\_educated[T.True] & 0.169 & 0.092 & 1.838 & 0.067 & [-0.012, 0.349] \\
occupation\_full\_time[T.True] & -0.061 & 0.092 & -0.666 & 0.506 & [-0.243, 0.120] \\
party\_Independent[T.True] & 0.114 & 0.108 & 1.057 & 0.291 & [-0.098, 0.326] \\
party\_Republican[T.True] & 0.053 & 0.107 & 0.498 & 0.619 & [-0.156, 0.262] \\
\bottomrule
\end{tabular}
\caption{Change in answers between pre- and post-study questionnaire on the question: \emph{``AI suggestions can make it more likely for me to participate in online discussions.''}}
\end{table}

\begin{table}[htp]
\centering
\begin{tabular}{lrrrrr}
\toprule
Variable & Coef. & Std. Err. & t & p-value & 95\% CI \\
\midrule
Intercept & 0.132 & 0.121 & 1.089 & 0.277 & [-0.106, 0.370] \\
treatment\_chat[T.True] & 0.139 & 0.120 & 1.156 & 0.248 & [-0.097, 0.376] \\
treatment\_conversation[T.True] & 0.094 & 0.121 & 0.777 & 0.437 & [-0.143, 0.331] \\
treatment\_feedback[T.True] & 0.068 & 0.121 & 0.559 & 0.576 & [-0.170, 0.305] \\
treatment\_suggestions[T.True] & 0.249 & 0.122 & 2.041 & 0.042 & [0.009, 0.488] \\
age\_old[T.True] & 0.166 & 0.078 & 2.129 & 0.034 & [0.013, 0.318] \\
gender\_not\_female[T.True] & -0.012 & 0.077 & -0.153 & 0.879 & [-0.163, 0.140] \\
education\_not\_educated[T.True] & 0.180 & 0.080 & 2.248 & 0.025 & [0.023, 0.336] \\
occupation\_full\_time[T.True] & 0.027 & 0.080 & 0.331 & 0.741 & [-0.131, 0.184] \\
party\_Independent[T.True] & -0.095 & 0.094 & -1.007 & 0.314 & [-0.279, 0.090] \\
party\_Republican[T.True] & -0.116 & 0.093 & -1.251 & 0.211 & [-0.298, 0.066] \\
\bottomrule
\end{tabular}
\caption{Change in answers between pre- and post-study questionnaire on the question: \emph{``AI can make online discussions more positive and less toxic.''}}
\end{table}

\begin{table}[htp]
\centering
\begin{tabular}{lrrrrr}
\toprule
Variable & Coef. & Std. Err. & t & p-value & 95\% CI \\
\midrule
Intercept & 0.226 & 0.126 & 1.800 & 0.072 & [-0.021, 0.473] \\
treatment\_chat[T.True] & -0.007 & 0.125 & -0.055 & 0.956 & [-0.252, 0.238] \\
treatment\_conversation[T.True] & -0.125 & 0.125 & -0.995 & 0.320 & [-0.371, 0.121] \\
treatment\_feedback[T.True] & 0.080 & 0.125 & 0.634 & 0.527 & [-0.167, 0.326] \\
treatment\_suggestions[T.True] & 0.229 & 0.126 & 1.812 & 0.070 & [-0.019, 0.477] \\
age\_old[T.True] & 0.080 & 0.080 & 0.993 & 0.321 & [-0.078, 0.238] \\
gender\_not\_female[T.True] & -0.120 & 0.080 & -1.497 & 0.135 & [-0.277, 0.037] \\
education\_not\_educated[T.True] & 0.089 & 0.083 & 1.076 & 0.282 & [-0.073, 0.252] \\
occupation\_full\_time[T.True] & -0.000 & 0.083 & -0.005 & 0.996 & [-0.164, 0.163] \\
party\_Independent[T.True] & 0.053 & 0.097 & 0.547 & 0.585 & [-0.138, 0.245] \\
party\_Republican[T.True] & -0.028 & 0.096 & -0.297 & 0.767 & [-0.217, 0.160] \\
\bottomrule
\end{tabular}
\caption{Change in answers between pre- and post-study questionnaire on the question: \emph{``AI can make discussions less polarizing.''}}
\end{table}

\begin{table}[htp]
\centering
\begin{tabular}{lrrrrr}
\toprule
Variable & Coef. & Std. Err. & t & p-value & 95\% CI \\
\midrule
Intercept & -0.015 & 0.125 & -0.117 & 0.907 & [-0.260, 0.231] \\
treatment\_chat[T.True] & 0.219 & 0.124 & 1.766 & 0.078 & [-0.024, 0.462] \\
treatment\_conversation[T.True] & -0.221 & 0.124 & -1.783 & 0.075 & [-0.465, 0.022] \\
treatment\_feedback[T.True] & 0.028 & 0.124 & 0.222 & 0.824 & [-0.217, 0.272] \\
treatment\_suggestions[T.True] & 0.075 & 0.125 & 0.596 & 0.551 & [-0.171, 0.320] \\
age\_old[T.True] & 0.096 & 0.080 & 1.201 & 0.230 & [-0.061, 0.253] \\
gender\_not\_female[T.True] & -0.103 & 0.079 & -1.304 & 0.193 & [-0.259, 0.052] \\
education\_not\_educated[T.True] & 0.082 & 0.082 & 1.006 & 0.315 & [-0.079, 0.243] \\
occupation\_full\_time[T.True] & 0.046 & 0.083 & 0.559 & 0.576 & [-0.116, 0.209] \\
party\_Independent[T.True] & 0.006 & 0.096 & 0.064 & 0.949 & [-0.183, 0.196] \\
party\_Republican[T.True] & 0.013 & 0.095 & 0.134 & 0.893 & [-0.174, 0.200] \\
\bottomrule
\end{tabular}
\caption{Change in answers between pre- and post-study questionnaire on the question: \emph{``AI can help reduce misinformation on social media.''}}
\end{table}

\begin{table}[htp]
\centering
\begin{tabular}{lrrrrr}
\toprule
Variable & Coef. & Std. Err. & t & p-value & 95\% CI \\
\midrule
Intercept & 0.273 & 0.104 & 2.617 & 0.009 & [0.068, 0.477] \\
treatment\_chat[T.True] & 0.261 & 0.103 & 2.536 & 0.011 & [0.059, 0.463] \\
treatment\_conversation[T.True] & 0.046 & 0.103 & 0.444 & 0.657 & [-0.157, 0.248] \\
treatment\_feedback[T.True] & 0.075 & 0.103 & 0.723 & 0.470 & [-0.128, 0.278] \\
treatment\_suggestions[T.True] & 0.185 & 0.105 & 1.770 & 0.077 & [-0.020, 0.390] \\
age\_old[T.True] & -0.092 & 0.066 & -1.381 & 0.168 & [-0.222, 0.039] \\
gender\_not\_female[T.True] & -0.073 & 0.066 & -1.104 & 0.270 & [-0.203, 0.057] \\
education\_not\_educated[T.True] & 0.117 & 0.068 & 1.705 & 0.089 & [-0.018, 0.251] \\
occupation\_full\_time[T.True] & -0.091 & 0.069 & -1.325 & 0.186 & [-0.227, 0.044] \\
party\_Independent[T.True] & -0.133 & 0.080 & -1.647 & 0.100 & [-0.290, 0.025] \\
party\_Republican[T.True] & -0.070 & 0.079 & -0.877 & 0.381 & [-0.225, 0.086] \\
\bottomrule
\end{tabular}
\caption{Change in answers between pre- and post-study questionnaire on the question: \emph{``AI-generated content is accurate and reliable.''}}
\end{table}

\begin{table}[htp]
\centering
\begin{tabular}{lrrrrr}
\toprule
Variable & Coef. & Std. Err. & t & p-value & 95\% CI \\
\midrule
Intercept & -0.275 & 0.124 & -2.208 & 0.028 & [-0.519, -0.030] \\
treatment\_chat[T.True] & -0.003 & 0.123 & -0.022 & 0.982 & [-0.244, 0.238] \\
treatment\_conversation[T.True] & 0.015 & 0.123 & 0.119 & 0.906 & [-0.227, 0.256] \\
treatment\_feedback[T.True] & 0.088 & 0.124 & 0.714 & 0.475 & [-0.154, 0.331] \\
treatment\_suggestions[T.True] & -0.082 & 0.125 & -0.656 & 0.512 & [-0.327, 0.163] \\
age\_old[T.True] & 0.048 & 0.079 & 0.610 & 0.542 & [-0.107, 0.204] \\
gender\_not\_female[T.True] & 0.043 & 0.079 & 0.552 & 0.581 & [-0.111, 0.198] \\
education\_not\_educated[T.True] & 0.105 & 0.081 & 1.291 & 0.197 & [-0.055, 0.265] \\
occupation\_full\_time[T.True] & 0.052 & 0.082 & 0.629 & 0.530 & [-0.110, 0.213] \\
party\_Independent[T.True] & 0.228 & 0.096 & 2.382 & 0.018 & [0.040, 0.417] \\
party\_Republican[T.True] & 0.139 & 0.094 & 1.468 & 0.143 & [-0.047, 0.324] \\
\bottomrule
\end{tabular}
\caption{Change in answers between pre- and post-study questionnaire on the question: \emph{``AI should be regulated to prevent misuse and ensure ethical use.''}}
\end{table}

\FloatBarrier

\subsection*{Participation Equality (Normalized Shannon Entropy)}

To measure the diversity of contributions within a round, we convert the counts of comments into probabilities ($p_i$) and compute the raw Shannon entropy:

$$
H=-\sum(p_i*\log(p_i))
$$

We normalize the entropy to the number of participants. The normalized Shannon entropy is an intuitive and effective measure to quantify how evenly contributions are distributed within each discussion. It ranges between 0 and 1, with 0 indicating complete contribution by a single participant and 1 representing a perfectly even distribution of contributions among participants. Based on the entropy values in each treatment group, we use bootstrapping to obtain confidence intervals and two-sided t-test for comparing treatments against control.

\section*{Supplementary Figures}
\label{sec:figures}

The figures below are based on questionnaire responses from 680 participants, distributed across 136 groups. This corresponds to 135 participants in each of the \textit{Control}, \textit{Suggestions}, \textit{Chat}, and \textit{Feedback} conditions, and 140 in the \textit{Conversation Starter} condition.

\subsection*{Pre-Study Questionnaire Answer Distributions}

We plot the distribution of answers to the pre-study questionnaire.

\subsubsection*{Demographics}

\begin{figure}[H]
    \centering
    \begin{minipage}[t]{0.48\textwidth}
        \centering
        \includegraphics[width=\linewidth]{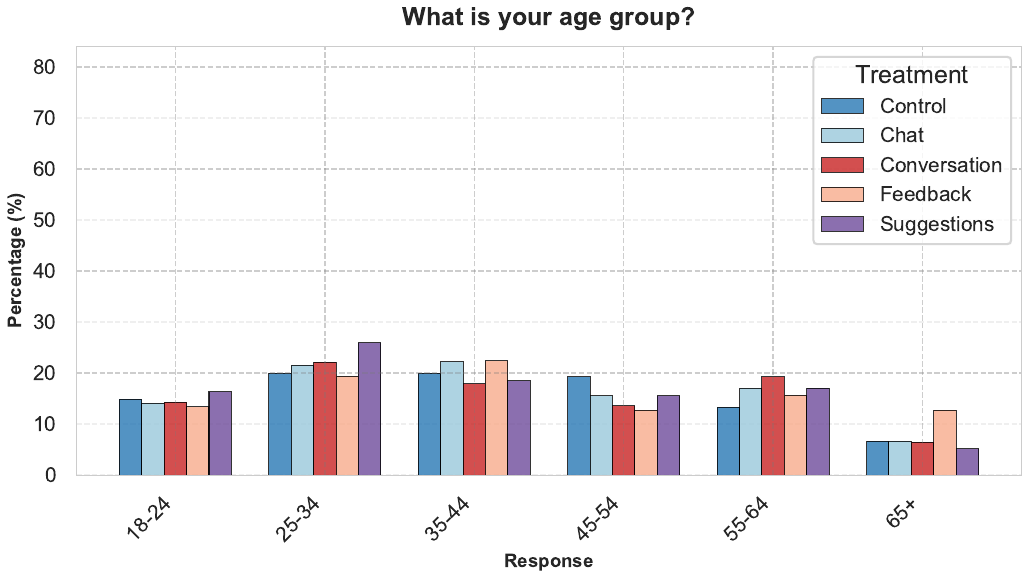}
    \end{minipage}%
    \hfill
    \begin{minipage}[t]{0.48\textwidth}
        \centering
        \includegraphics[width=\linewidth]{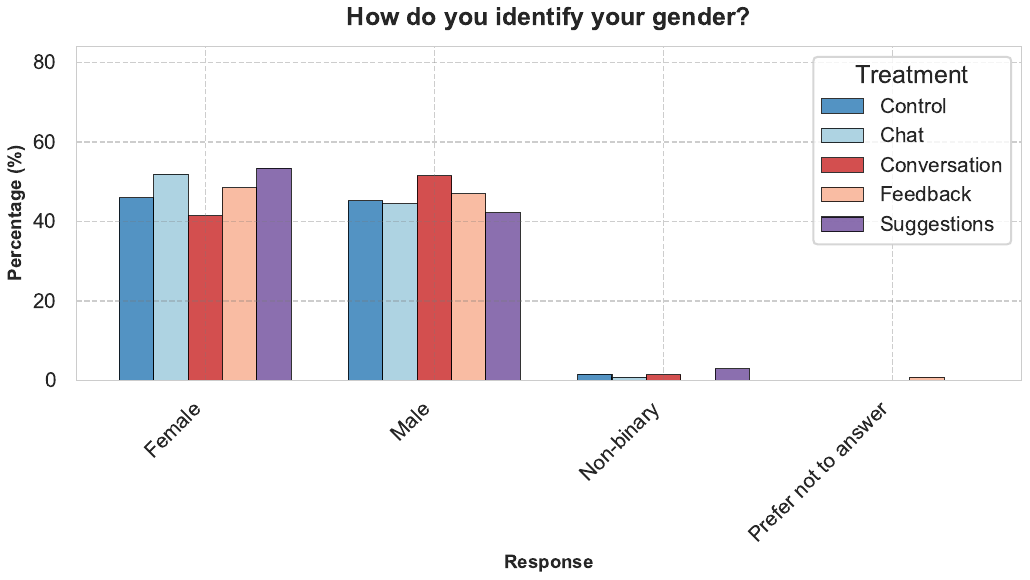}
    \end{minipage}
    
    \vspace{0.5cm}
    
    \begin{minipage}[t]{0.48\textwidth}
        \centering
        \includegraphics[width=\linewidth]{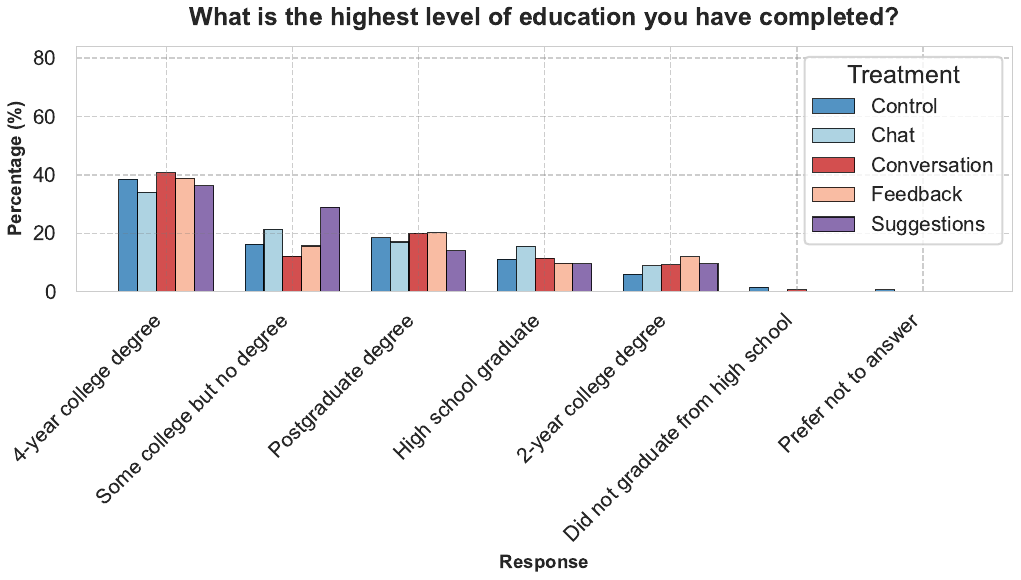}
    \end{minipage}%
    \hfill
    \begin{minipage}[t]{0.48\textwidth}
        \centering
        \includegraphics[width=\linewidth]{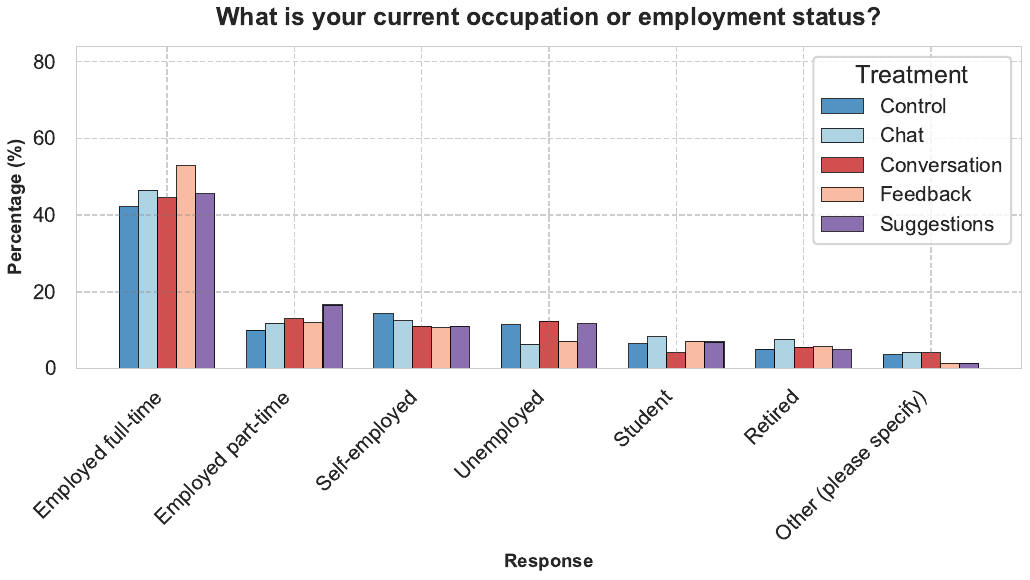}
    \end{minipage}
\end{figure}

\begin{figure}[H]
    \centering    

    \begin{minipage}[t]{0.48\textwidth}
        \centering
        \includegraphics[width=\linewidth]{figures/initial-questionnaire/demographicInfo_occupation.pdf}
    \end{minipage}%
    \hfill
    \begin{minipage}[t]{0.48\textwidth}
        \centering
        \includegraphics[width=\linewidth]{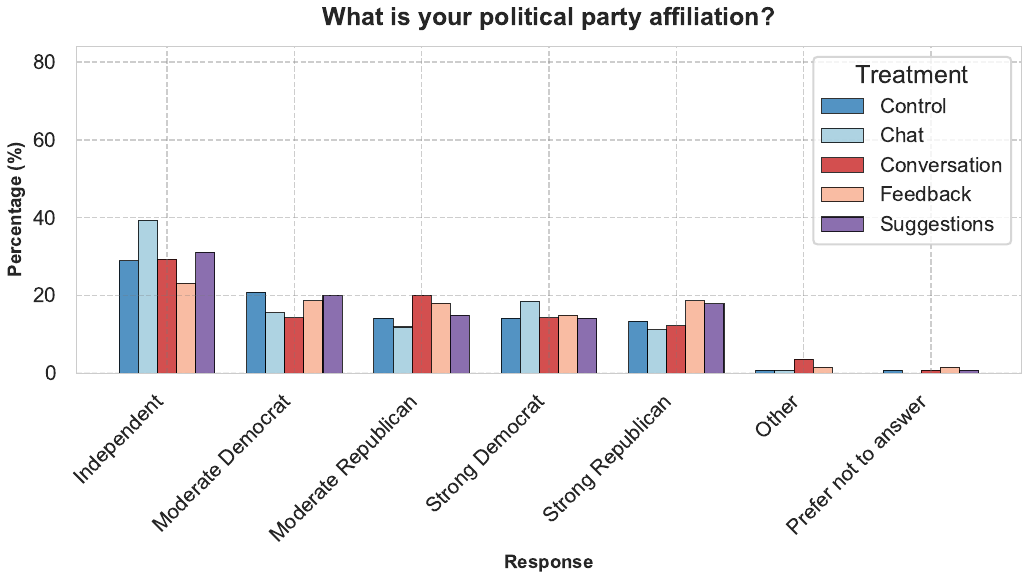}
    \end{minipage}

    \label{fig:prequestionnaire_distributions-demographics}
    \caption{[\textit{Continued from last page}] Pre-Study Questionnaire Answer Distribution: Demographics.}
\end{figure}

\subsubsection*{Social Media Questions}

\begin{figure}[H]
    \centering
    \begin{minipage}[t]{0.48\textwidth}
        \centering
        \includegraphics[width=\linewidth]{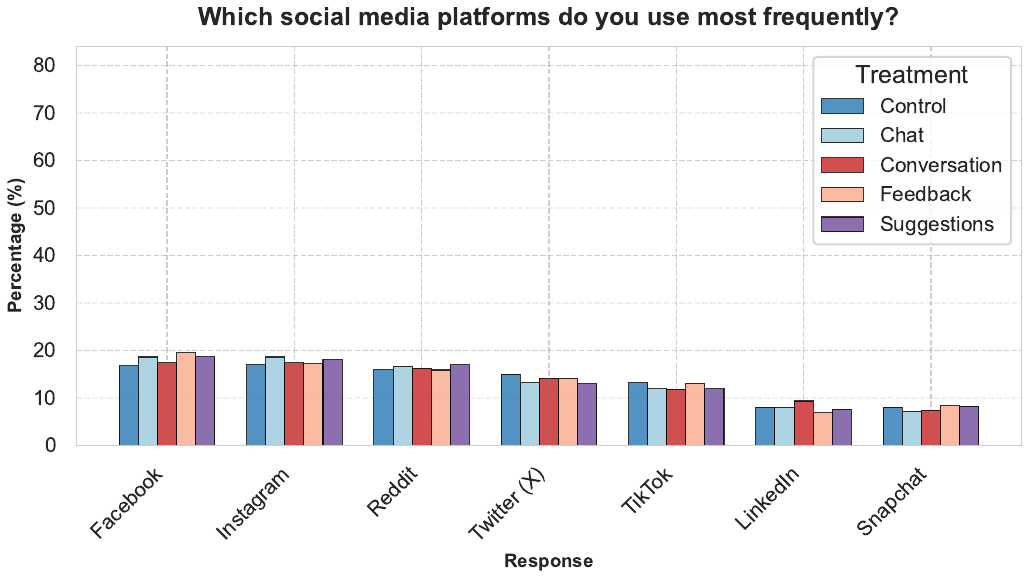}
    \end{minipage}%
    \hfill
    \begin{minipage}[t]{0.48\textwidth}
        \centering
        \includegraphics[width=\linewidth]{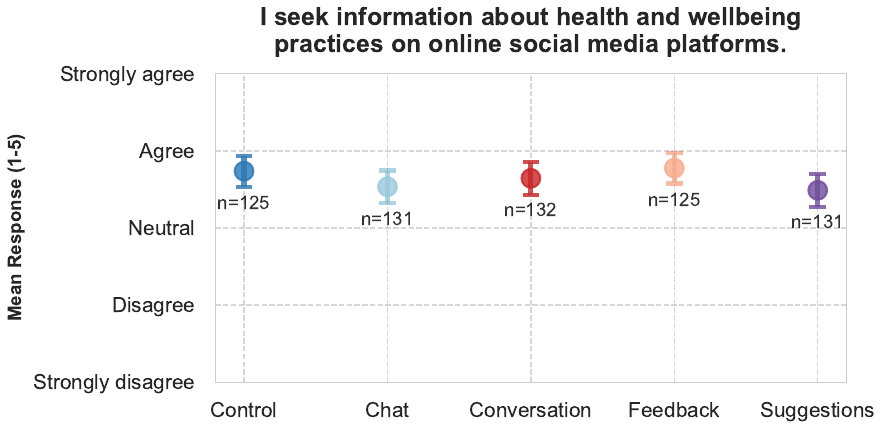}
    \end{minipage}

    \vspace{0.5cm} 

    \begin{minipage}[t]{0.48\textwidth}
        \centering
        \includegraphics[width=\linewidth]{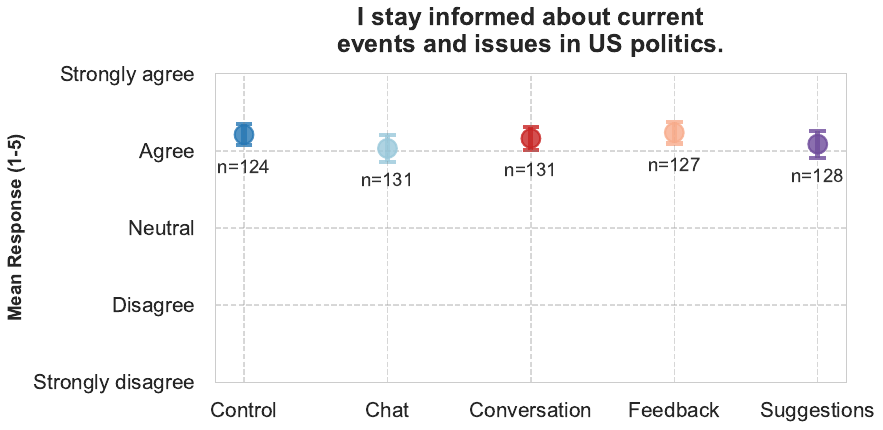}
    \end{minipage}%
    \hfill
    \begin{minipage}[t]{0.48\textwidth}
        \centering
        \includegraphics[width=\linewidth]{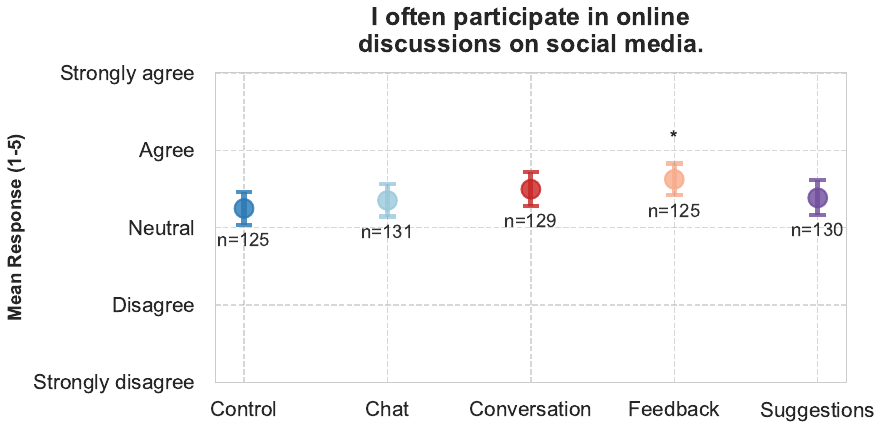}
    \end{minipage}
    
    \vspace{0.5cm} 
    
    \begin{minipage}[t]{0.48\textwidth}
        \centering
        \includegraphics[width=\linewidth]{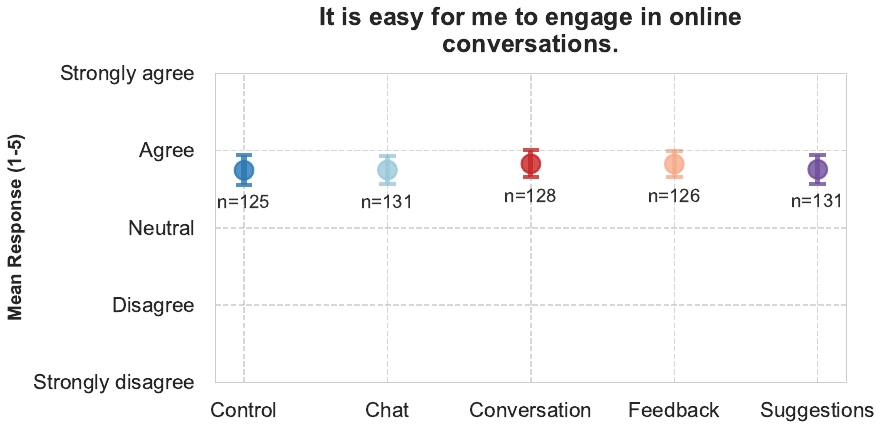}
    \end{minipage}%
    \hfill
    \begin{minipage}[t]{0.48\textwidth}
        \centering
        \includegraphics[width=\linewidth]{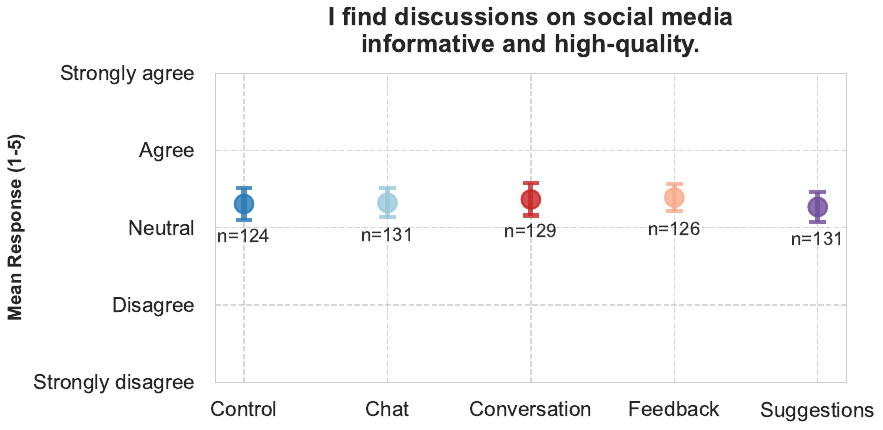}
    \end{minipage}
    \label{fig:prequestionnaire_distributions}
\end{figure}

\begin{figure}[H]
    \centering
    
    \begin{minipage}[t]{0.48\textwidth}
        \centering
        \includegraphics[width=\linewidth]{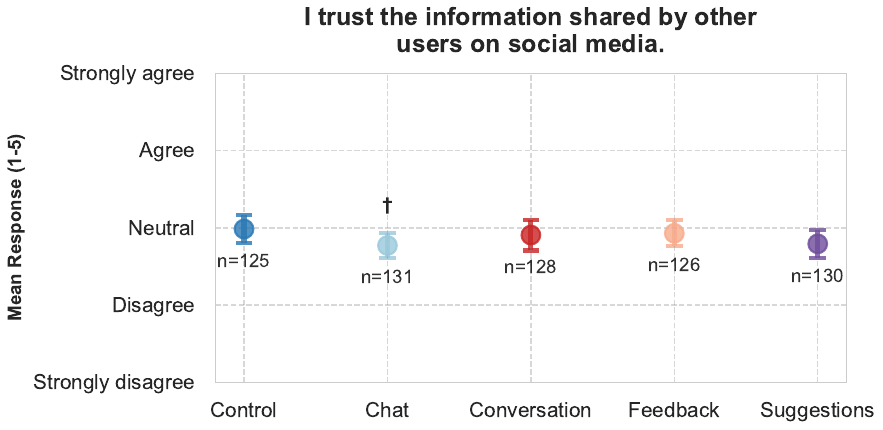}
    \end{minipage}%
    \hfill
    \begin{minipage}[t]{0.48\textwidth}
        \centering
        \includegraphics[width=\linewidth]{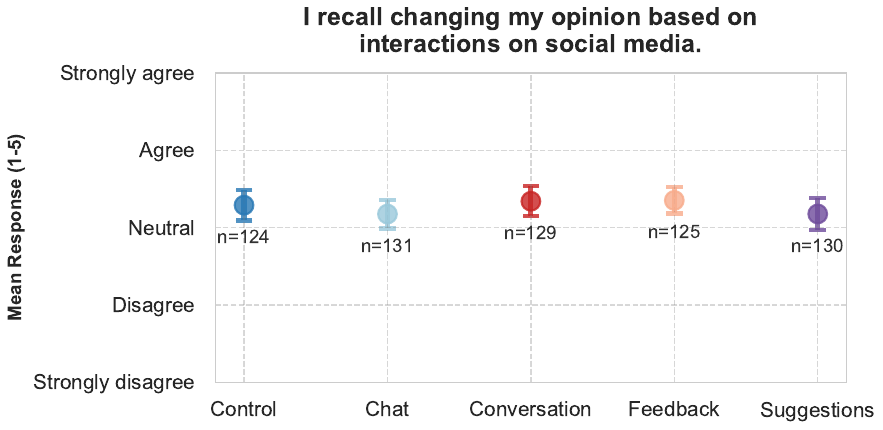}
    \end{minipage}

    \vspace{0.5cm} 
    
    \begin{minipage}[t]{0.48\textwidth}
        \centering
        \includegraphics[width=\linewidth]{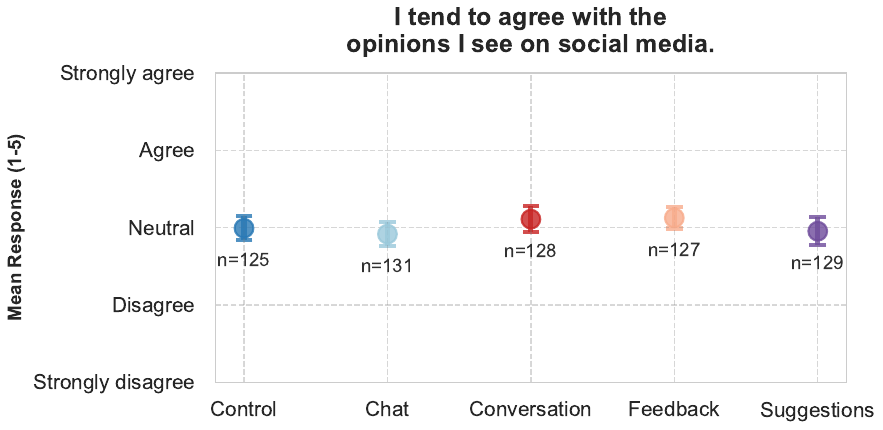}
    \end{minipage}%
    \hfill
    \begin{minipage}[t]{0.48\textwidth}
        \centering
        \includegraphics[width=\linewidth]{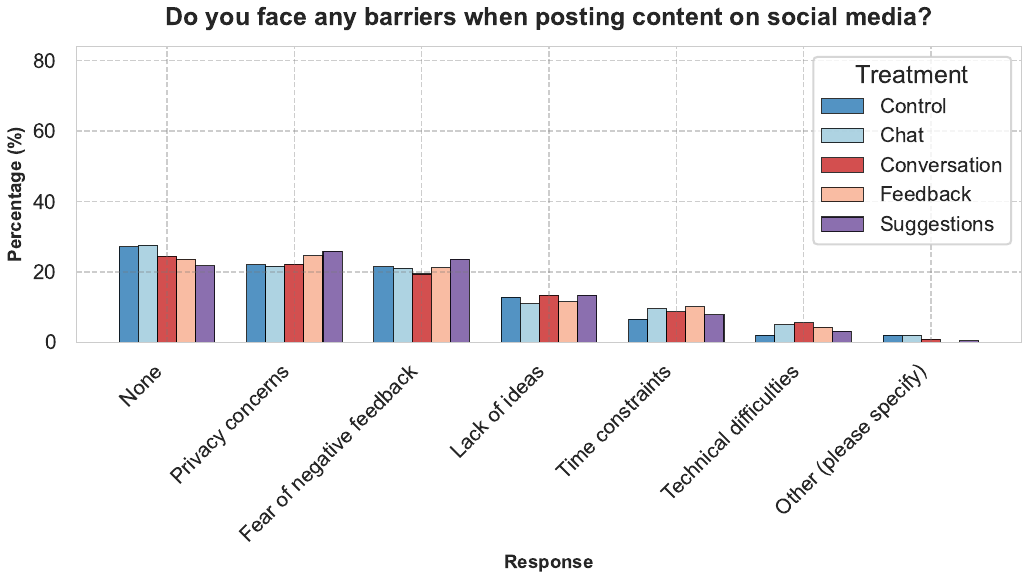}
    \end{minipage}

    \vspace{0.5cm} 
    
    \label{fig:prequestionnaire_distributions:some}
    \caption{[\textit{Continued from last page}] Pre-Study Questionnaire Answer Distribution: Questions related social media.}
\end{figure}

\subsubsection*{AI Related to Social Media Questions}

\begin{figure}[H]
    \centering
    \begin{minipage}[t]{0.48\textwidth}
        \centering
        \includegraphics[width=\linewidth]{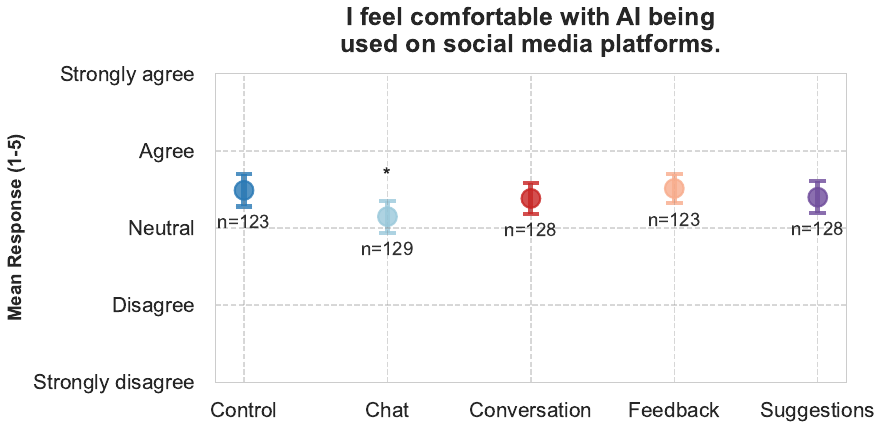}
    \end{minipage}%
    \hfill
    \begin{minipage}[t]{0.48\textwidth}
        \centering
        \includegraphics[width=\linewidth]{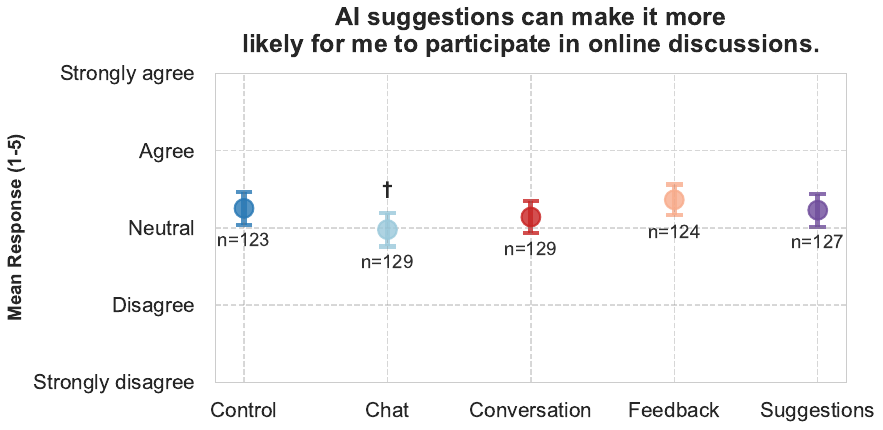}
    \end{minipage}
    \vspace{0.5cm} 
    \begin{minipage}[t]{0.48\textwidth}
        \centering
        \includegraphics[width=\linewidth]{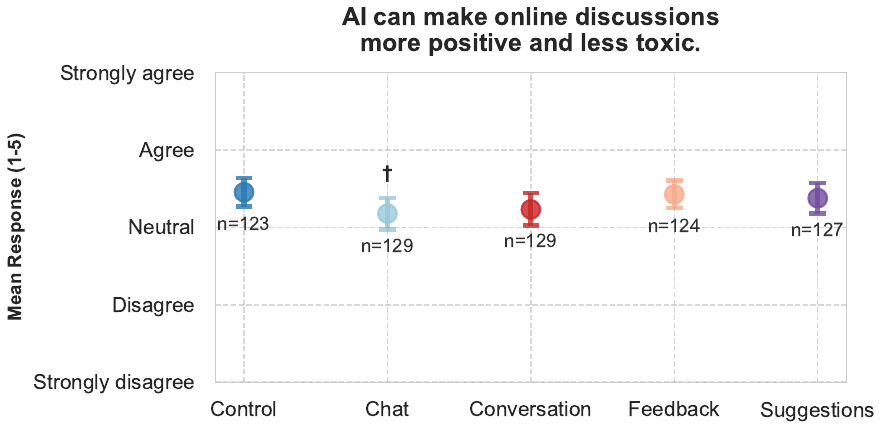}
    \end{minipage}%
    \hfill
    \begin{minipage}[t]{0.48\textwidth}
        \centering
        \includegraphics[width=\linewidth]{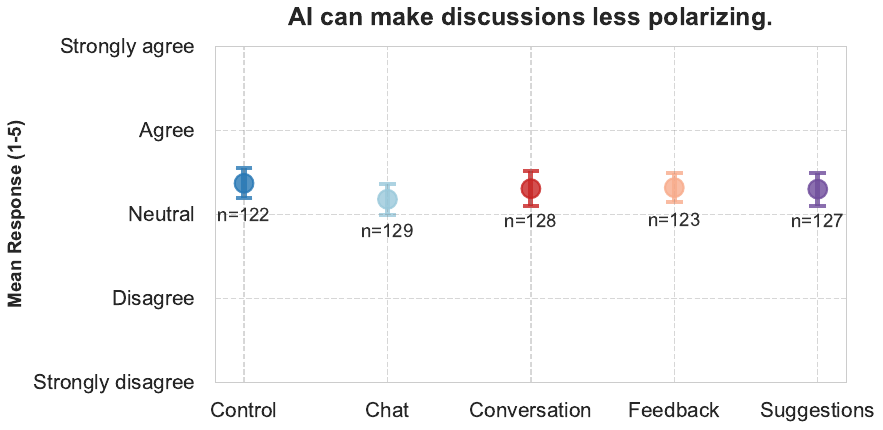}
    \end{minipage}
\end{figure}

\begin{figure}[H]
    \centering
    
    \begin{minipage}[t]{0.48\textwidth}
        \centering
        \includegraphics[width=\linewidth]{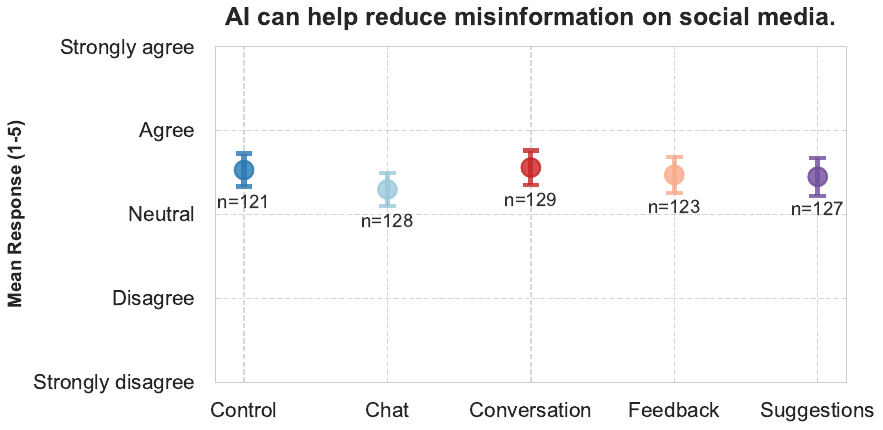}
    \end{minipage}%
    \hfill
    \begin{minipage}[t]{0.48\textwidth}
        \centering
        \includegraphics[width=\linewidth]{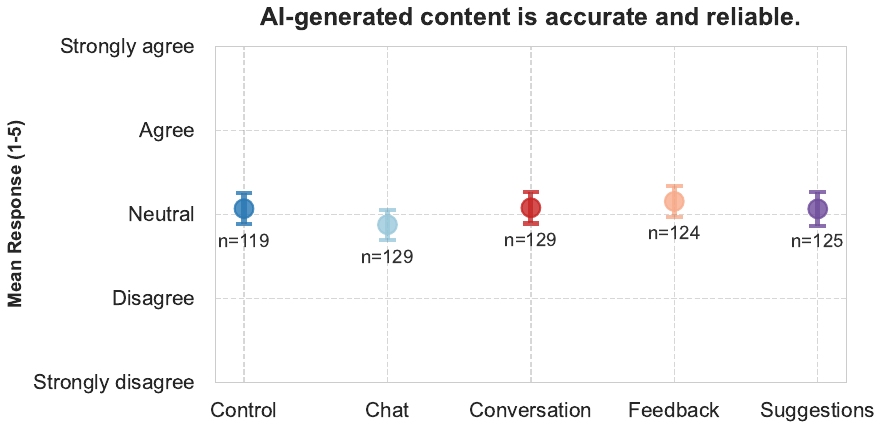}
    \end{minipage}
    \begin{minipage}[t]{0.48\textwidth}
        \centering
        \includegraphics[width=\linewidth]{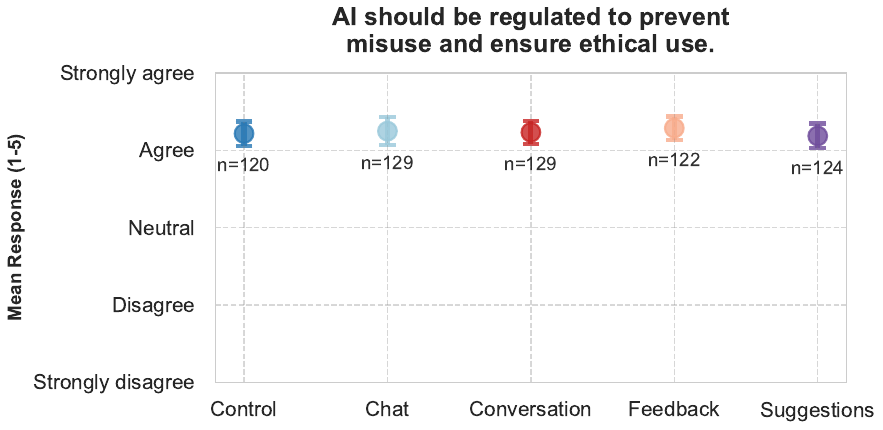}
    \end{minipage}%
    
    \label{fig:prequestionnaire_distributions-ai-some}
    \caption{[\textit{Continued from last page}] Pre-Study Questionnaire Answer Distribution: Questions on AI related to social media.}
\end{figure}

\subsection*{Post-Study Answer Distributions}

\subsubsection*{Platform Experience}

\begin{figure}[H]
    \centering
    \begin{minipage}[t]{0.48\textwidth}
        \centering
        \includegraphics[width=\linewidth]{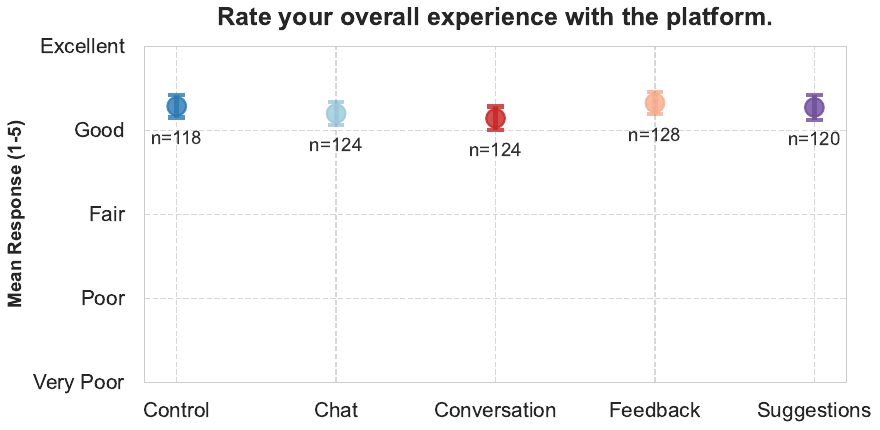}
    \end{minipage}%
    \hfill
    \begin{minipage}[t]{0.48\textwidth}
        \centering
        \includegraphics[width=\linewidth]{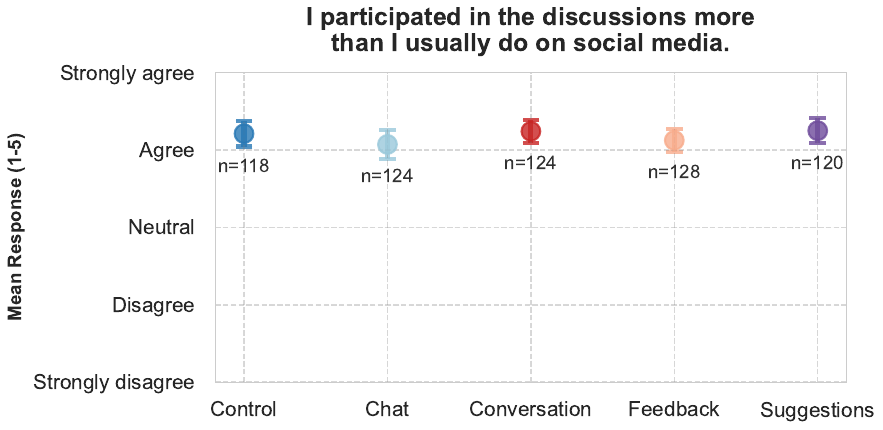}
    \end{minipage}
    
    \vspace{0.5cm} 
    
    \begin{minipage}[t]{0.48\textwidth}
        \centering
        \includegraphics[width=\linewidth]{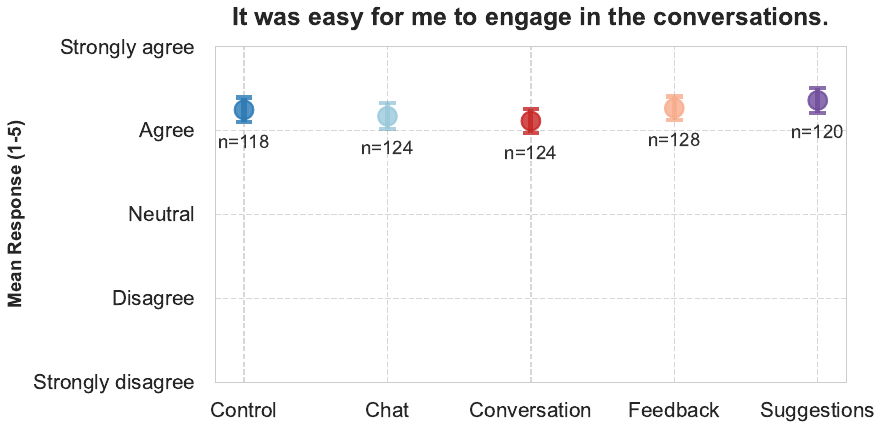}
    \end{minipage}%
    \hfill
    \begin{minipage}[t]{0.48\textwidth}
        \centering
        \includegraphics[width=\linewidth]{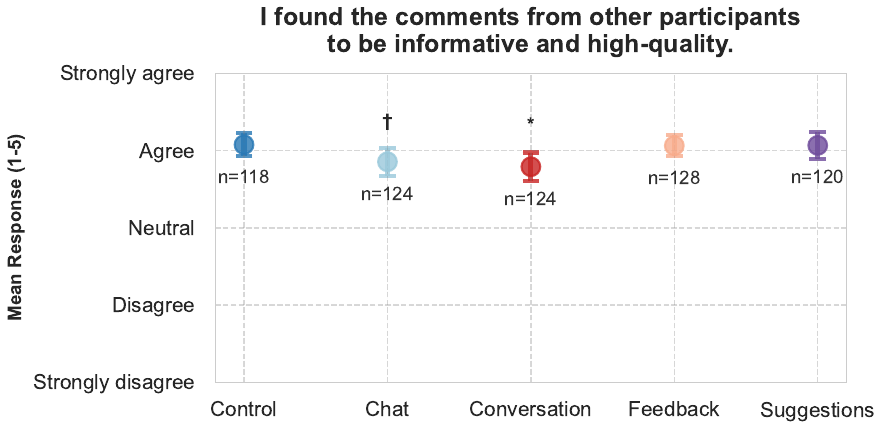}
    \end{minipage}
    
\end{figure}
\begin{figure}[H]
    \centering    

    \begin{minipage}[t]{0.48\textwidth}
        \centering
        \includegraphics[width=\linewidth]{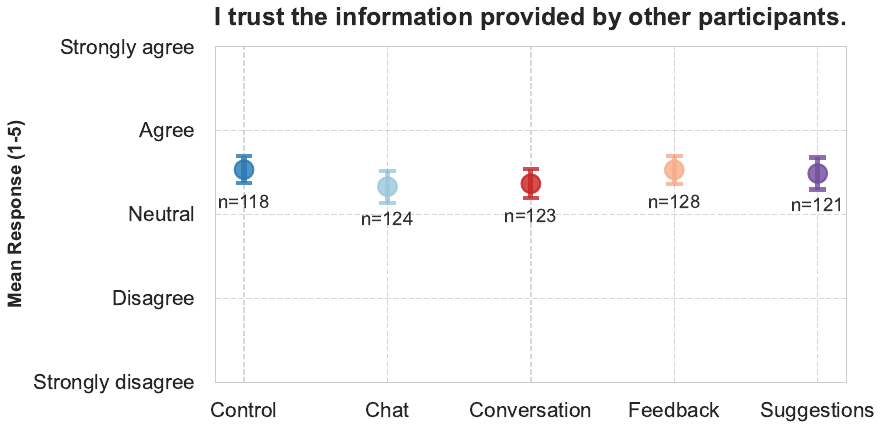}
    \end{minipage}%
    \hfill
    \begin{minipage}[t]{0.48\textwidth}
        \centering
        \includegraphics[width=\linewidth]{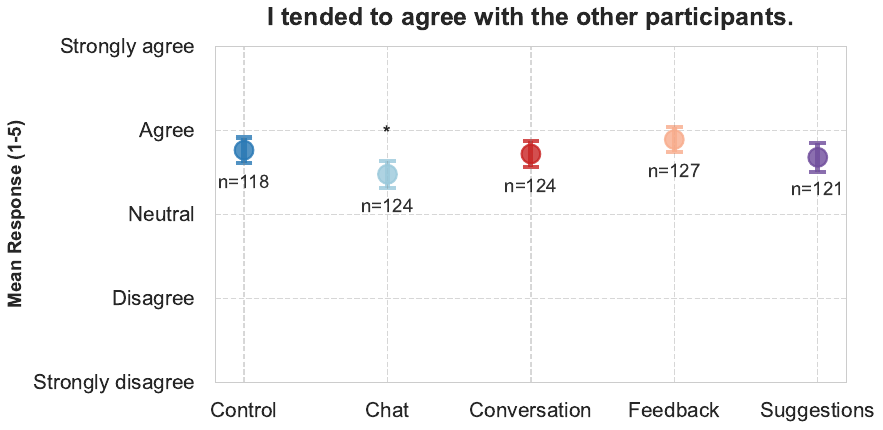}
    \end{minipage}

    \vspace{0.5cm} 
    
    \begin{minipage}[t]{0.48\textwidth}
        \centering
        \includegraphics[width=\linewidth]{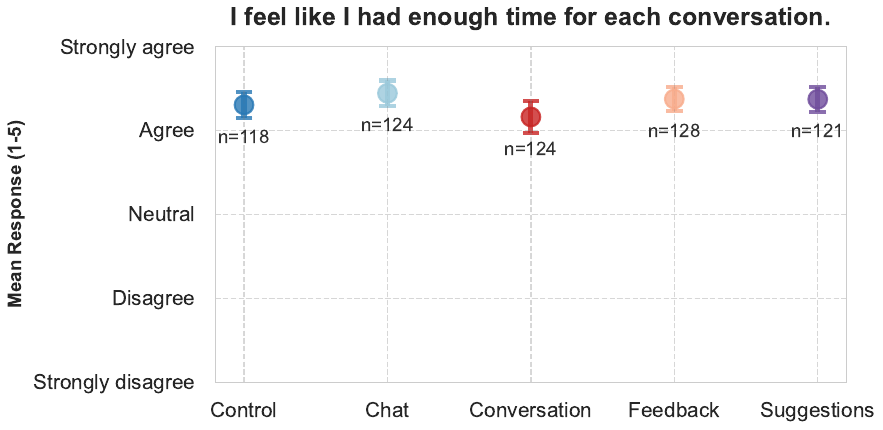}
    \end{minipage}%
    \hfill
    \begin{minipage}[t]{0.48\textwidth}
        \centering
        \includegraphics[width=\linewidth]{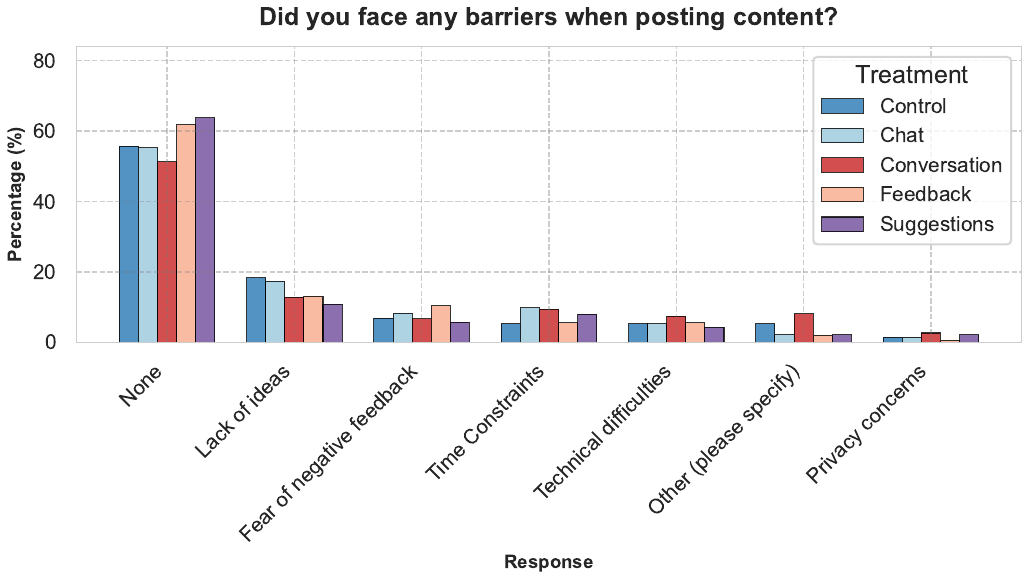}
    \end{minipage}

    \label{fig:postquestionnaire_distributions-platform}
    \caption{[\textit{Continued from last page}] Post-Study Questionnaire Answer Distribution: Questions on platform experience.}
    
\end{figure}

\subsubsection*{AI Evaluation (only for treatment participants)}

\begin{figure}[H]
    \centering
    \begin{minipage}[t]{0.48\textwidth}
        \centering
        \includegraphics[width=\linewidth]{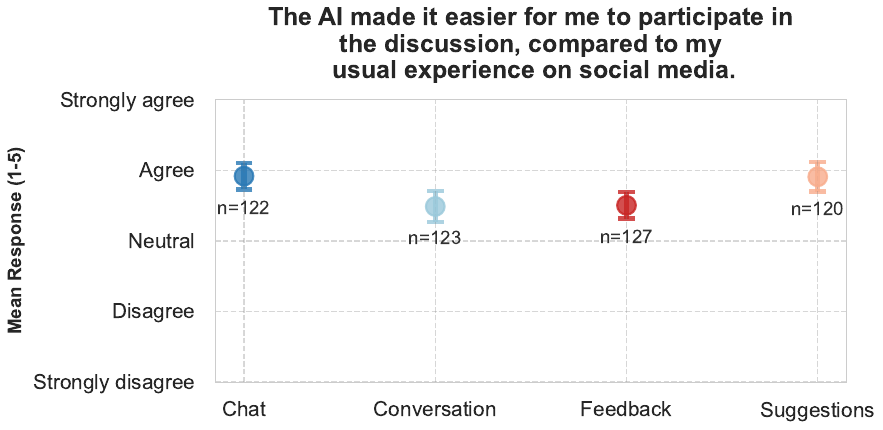}
    \end{minipage}%
    \hfill
    \begin{minipage}[t]{0.48\textwidth}
        \centering
        \includegraphics[width=\linewidth]{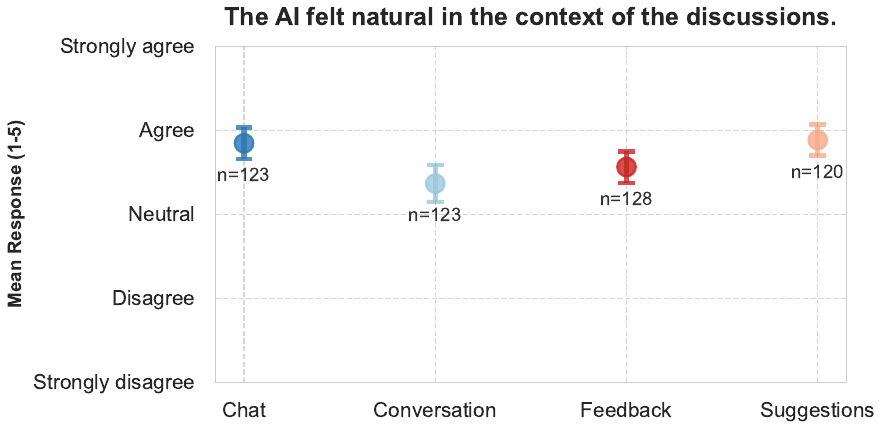}
    \end{minipage}
    
\end{figure}
\begin{figure}[H]
    
    \begin{minipage}[t]{0.48\textwidth}
        \centering
        \includegraphics[width=\linewidth]{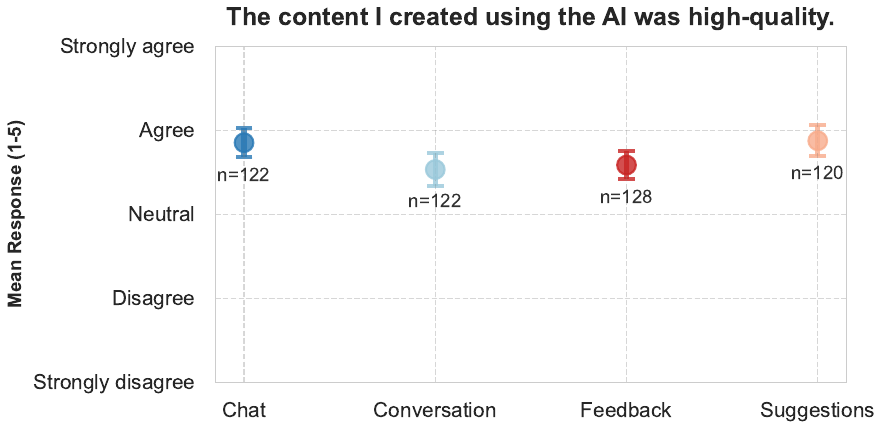}
    \end{minipage}%
    \hfill
    \begin{minipage}[t]{0.48\textwidth}
        \centering
        \includegraphics[width=\linewidth]{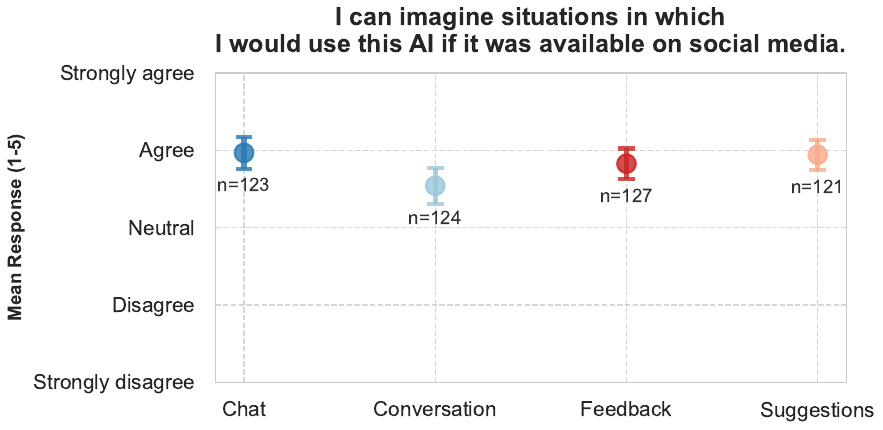}
    \end{minipage}

    \label{fig:postquestionnaire_distributions-ai-tool}
    \caption{Post-Study Questionnaire Answer Distribution: Questions on the AI tool. Only applicable for participants in the treatments groups (\textit{Chat}, \textit{Conversation Starter}, \textit{Feedback}, and \textit{Suggestions}).}
    
\end{figure}

\subsubsection*{AI Related to Social Media Questions}

\begin{figure}[H]
    \centering
    \begin{minipage}[t]{0.48\textwidth}
        \centering
        \includegraphics[width=\linewidth]{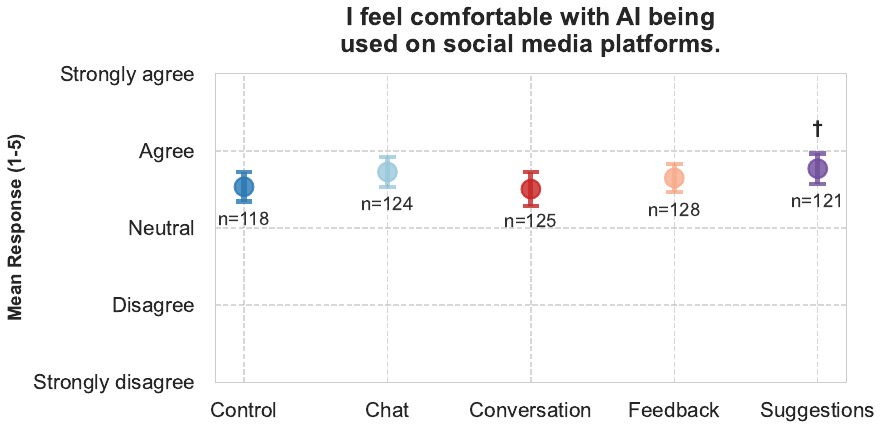}
    \end{minipage}%
    \hfill
    \begin{minipage}[t]{0.48\textwidth}
        \centering
        \includegraphics[width=\linewidth]{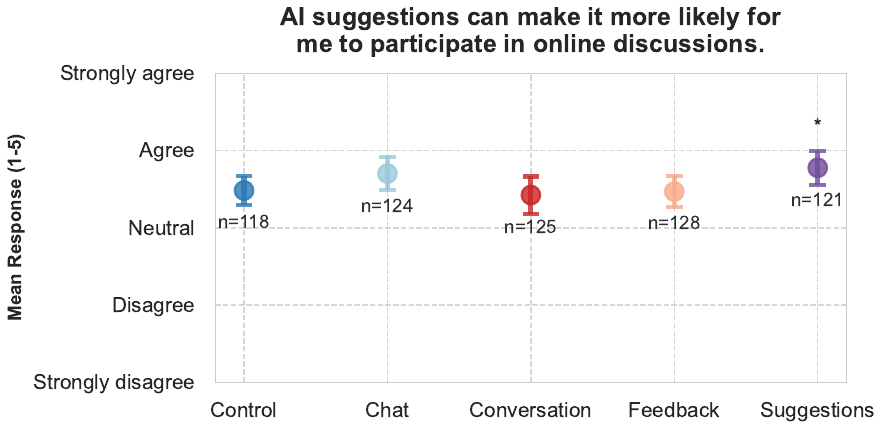}
    \end{minipage}

    \vspace{0.5cm} 

    \begin{minipage}[t]{0.48\textwidth}
        \centering
        \includegraphics[width=\linewidth]{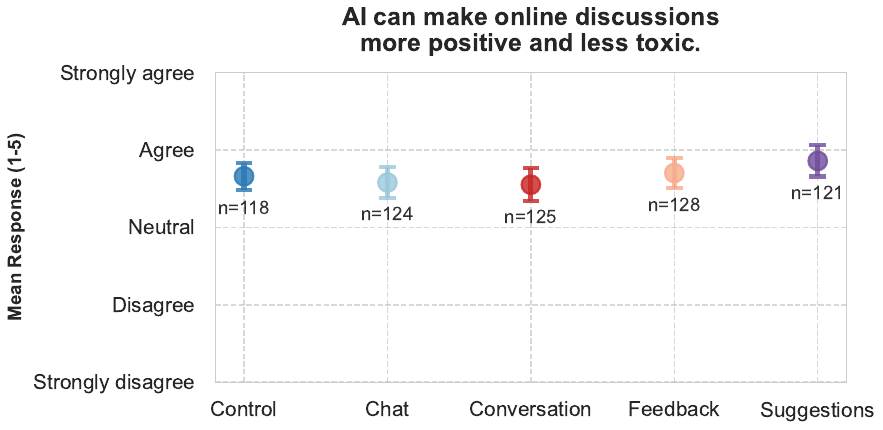}
    \end{minipage}%
    \hfill
    \begin{minipage}[t]{0.48\textwidth}
        \centering
        \includegraphics[width=\linewidth]{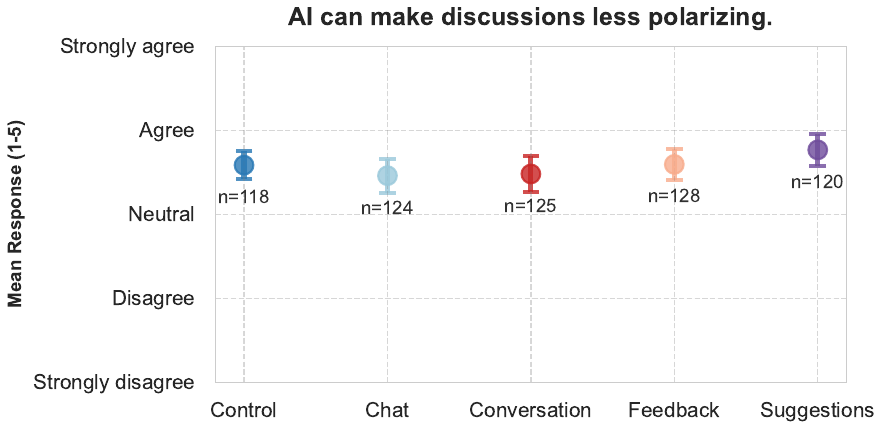}
    \end{minipage}

    \vspace{0.5cm} 

    \begin{minipage}[t]{0.48\textwidth}
        \centering
        \includegraphics[width=\linewidth]{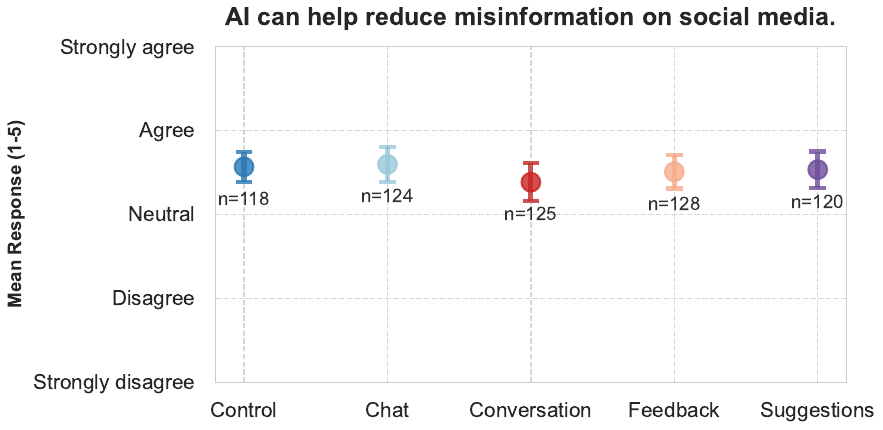}
    \end{minipage}%
    \hfill
    \begin{minipage}[t]{0.48\textwidth}
        \centering
        \includegraphics[width=\linewidth]{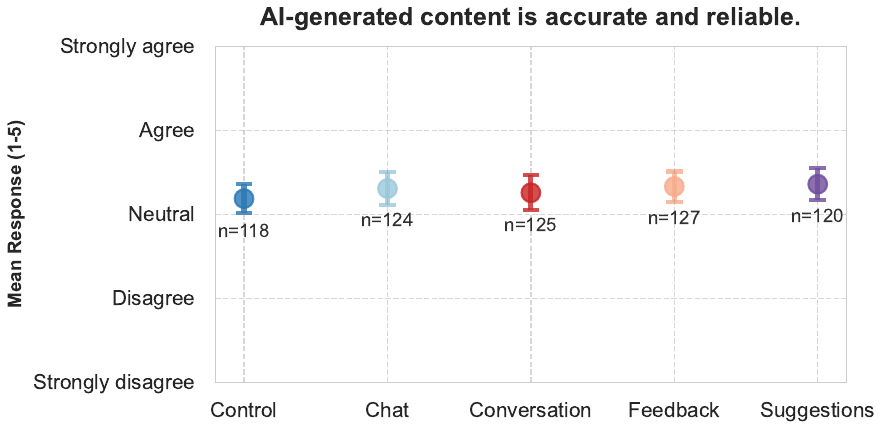}
    \end{minipage}
    
    \begin{minipage}[t]{0.48\textwidth}
        \centering
        \includegraphics[width=\linewidth]{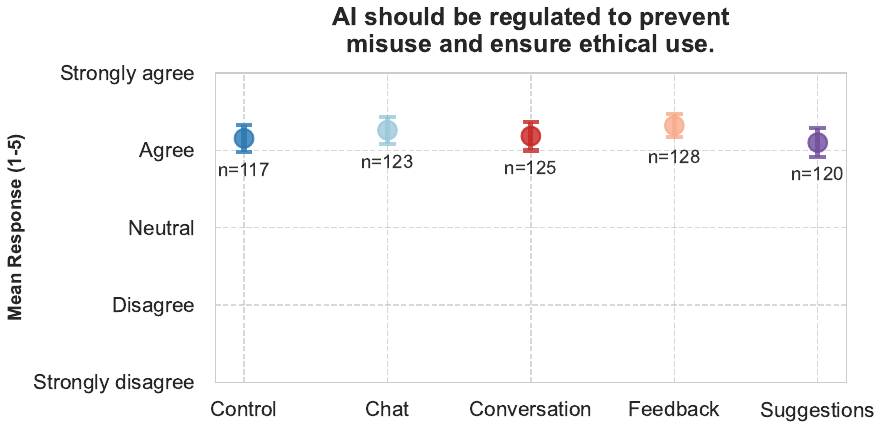}
    \end{minipage}%

    \label{fig:postquestionnaire_distributions-ai-sime}
    \caption{Post-Study Questionnaire Answer Distribution: Questions on AI in relation to social media.}
\end{figure}

\subsection*{Treatment Effects: Pre- and Post-Study Questionnaire Differences}

For the subset of questions included in both the pre- and post-study questionnaires, we assess the individual change during the study. The plots below display group-level mean differences and 95\% confidence intervals based on bootstrapping. Significance levels are based on two-sided \textit{t}-tests comparing each treatment group to the control.

\begin{figure}[H]
    \centering
    \begin{minipage}[t]{0.48\textwidth}
        \centering
        \includegraphics[width=\linewidth]{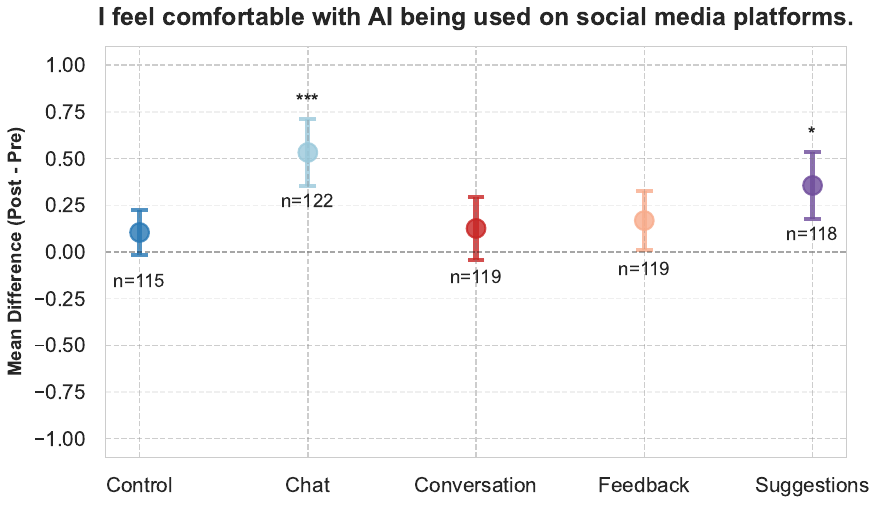}
    \end{minipage}%
    \hfill
    \begin{minipage}[t]{0.48\textwidth}
        \centering
        \includegraphics[width=\linewidth]{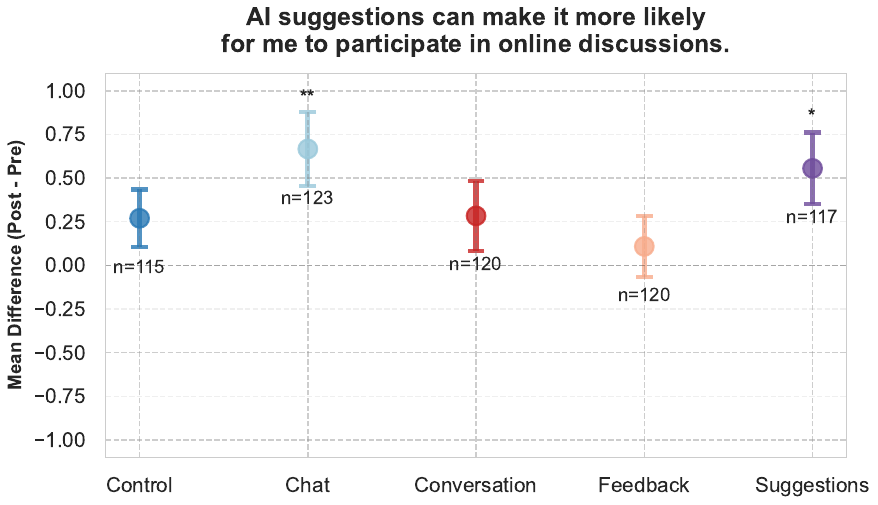}
    \end{minipage}

    \vspace{0.5cm} 

    \begin{minipage}[t]{0.48\textwidth}
        \centering
        \includegraphics[width=\linewidth]{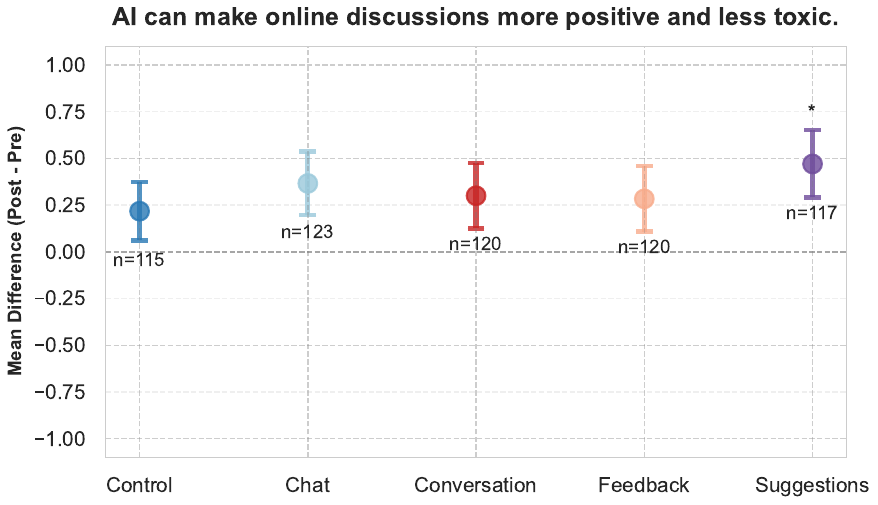}
    \end{minipage}%
    \hfill
    \begin{minipage}[t]{0.48\textwidth}
        \centering
        \includegraphics[width=\linewidth]{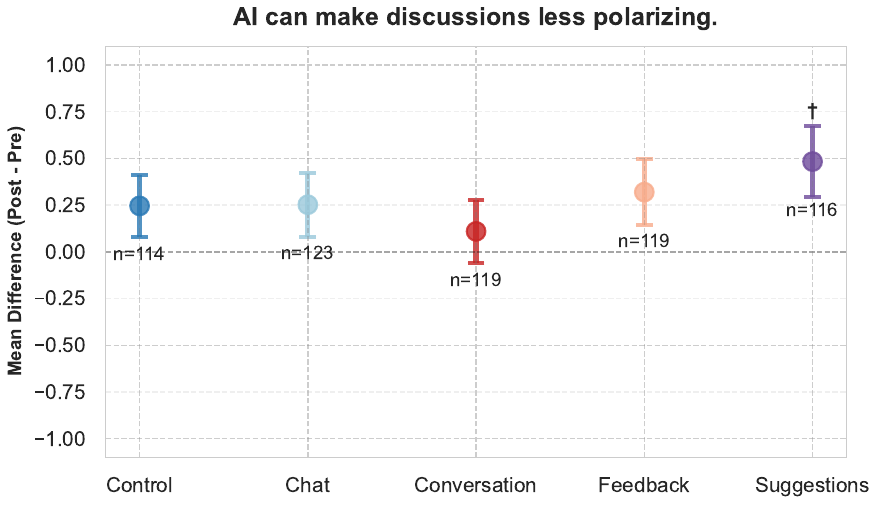}
    \end{minipage}
    
    \vspace{0.5cm} 
    
    \begin{minipage}[t]{0.48\textwidth}
        \centering
        \includegraphics[width=\linewidth]{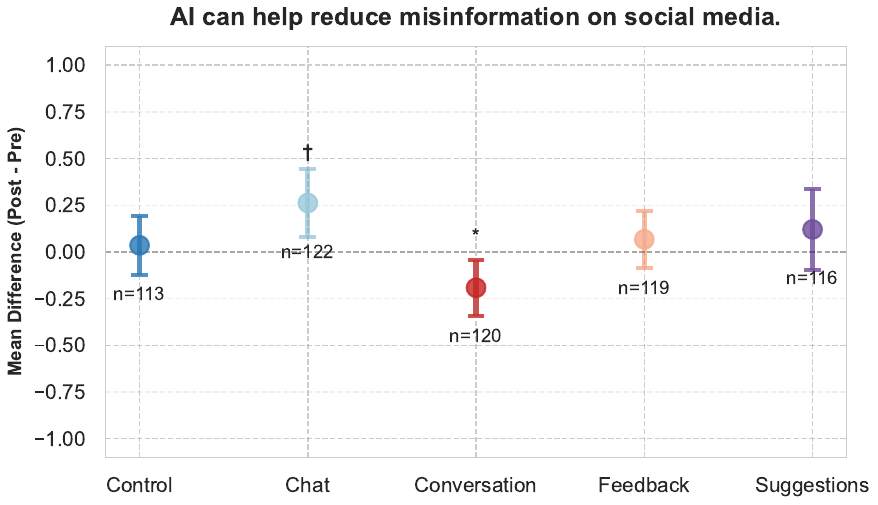}
    \end{minipage}%
    \hfill
    \begin{minipage}[t]{0.48\textwidth}
        \centering
        \includegraphics[width=\linewidth]{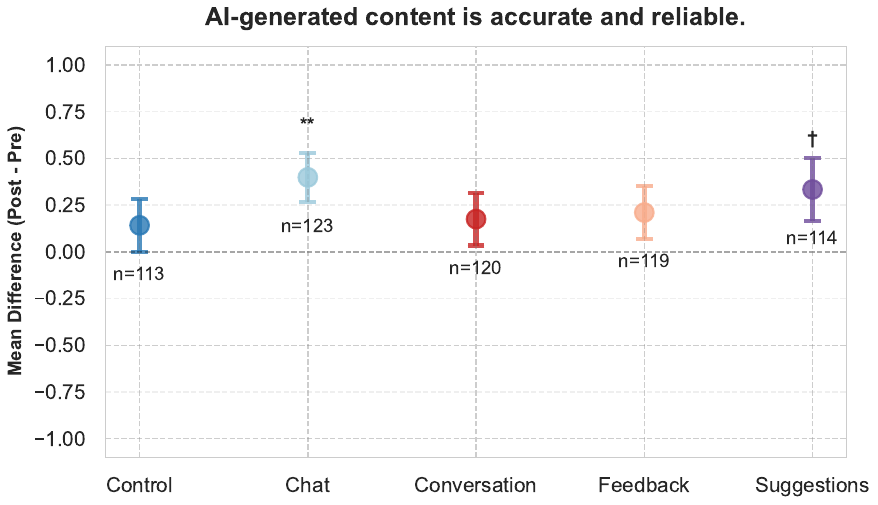}
    \end{minipage}

\end{figure}
\begin{figure}[H]
    \centering

    \begin{minipage}[t]{0.48\textwidth}
        \centering
        \includegraphics[width=\linewidth]{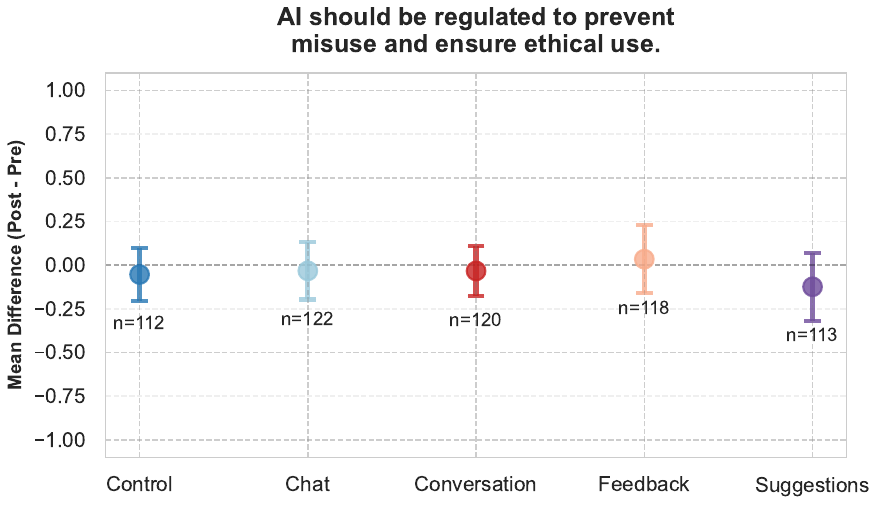}
    \end{minipage}%

    \label{fig:treatment-effects}
    \caption{Treatment Effects: Measuring $\Delta$ in answers between pre- and post-study questions.}
    
\end{figure}

\subsection*{Ratings of Users and Comments}

As a part of the post-study questionnaire, we ask users to rate comments received to their own comment. We randomly sample up to 10 comments. Additionally, we also ask users to rate other participants in their study-group. Comment and user ratings display individual user mean values with bootstrapped 95\% confidence interval and using two-sided t-test. For the questions with categorical options, we use permutation test.

\begin{figure}[H]
    \centering

    \begin{minipage}[t]{0.48\textwidth}
        \centering
        \includegraphics[width=\linewidth]{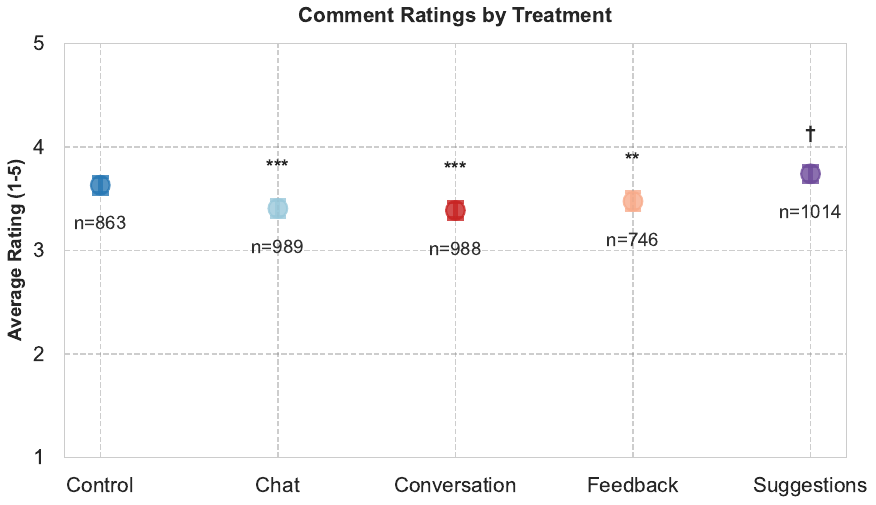}
    \end{minipage}%
    \hfill
    \begin{minipage}[t]{0.48\textwidth}
        \centering
        \includegraphics[width=\linewidth]{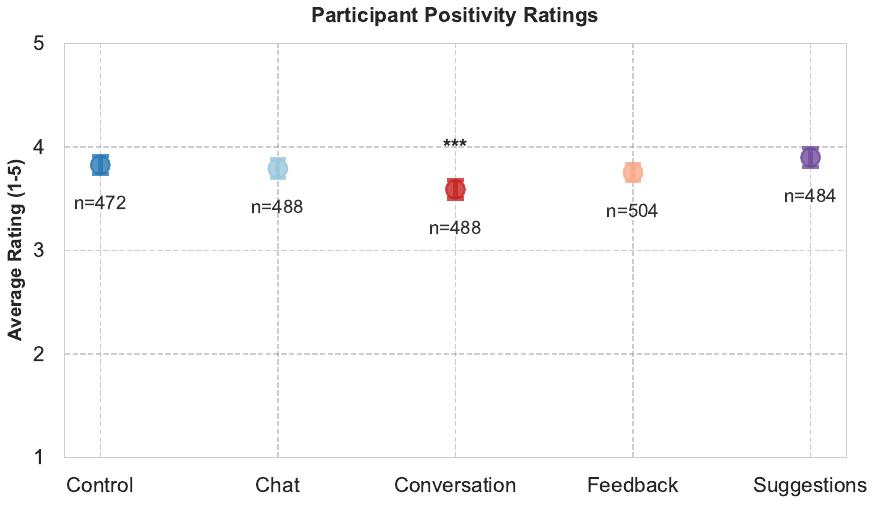}
    \end{minipage}

    \vspace{0.5cm} 
    
    \begin{minipage}[t]{0.48\textwidth}
        \centering
        \includegraphics[width=\linewidth]{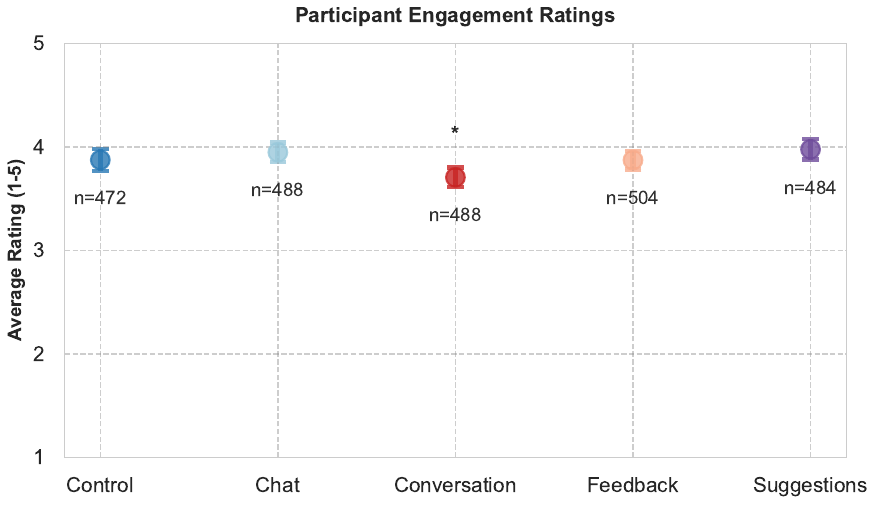}
    \end{minipage}%
    \hfill
    \begin{minipage}[t]{0.48\textwidth}
        \centering
        \includegraphics[width=\linewidth]{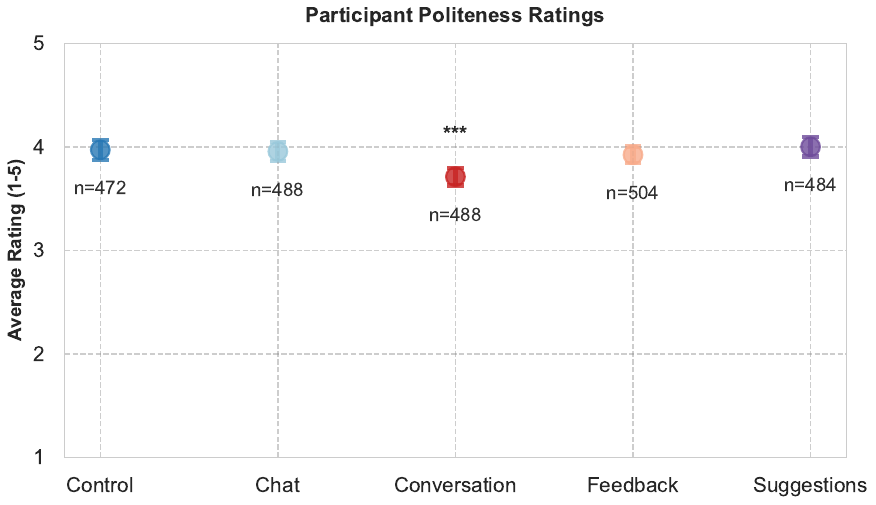}
    \end{minipage}
    \vspace{0.5cm} 
    \begin{minipage}[t]{0.48\textwidth}
        \centering
        \includegraphics[width=\linewidth]{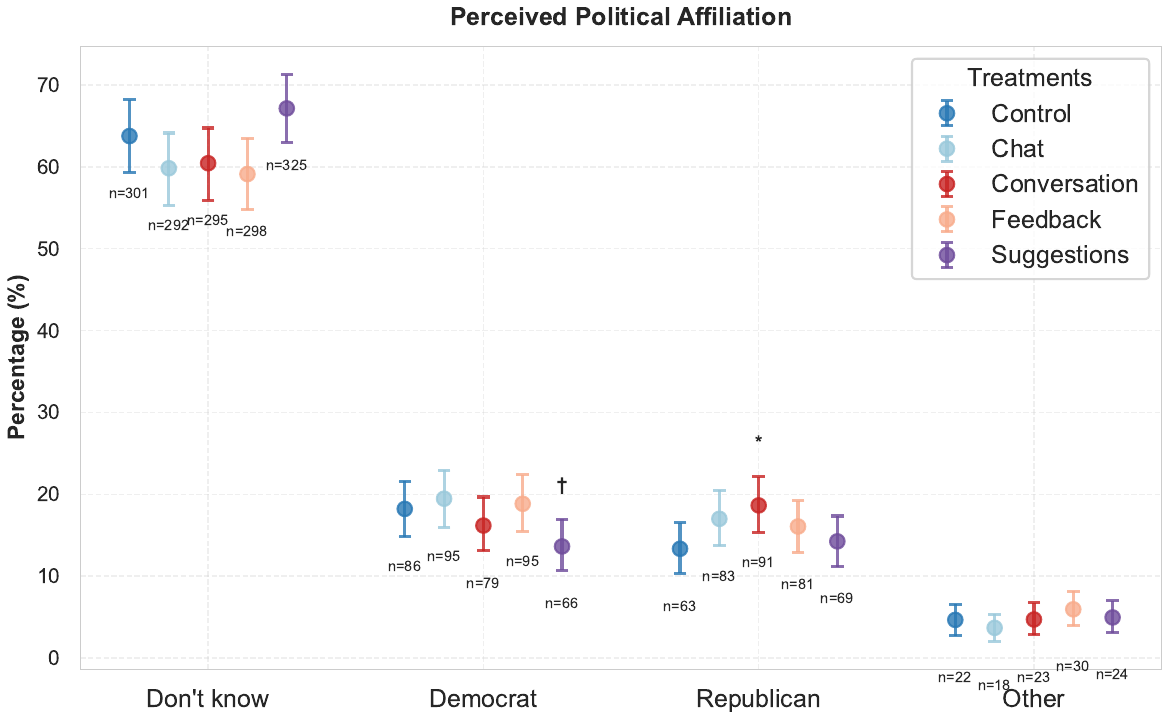}
    \end{minipage}%
    \hfill
    \begin{minipage}[t]{0.48\textwidth}
        \centering
        \includegraphics[width=\linewidth]{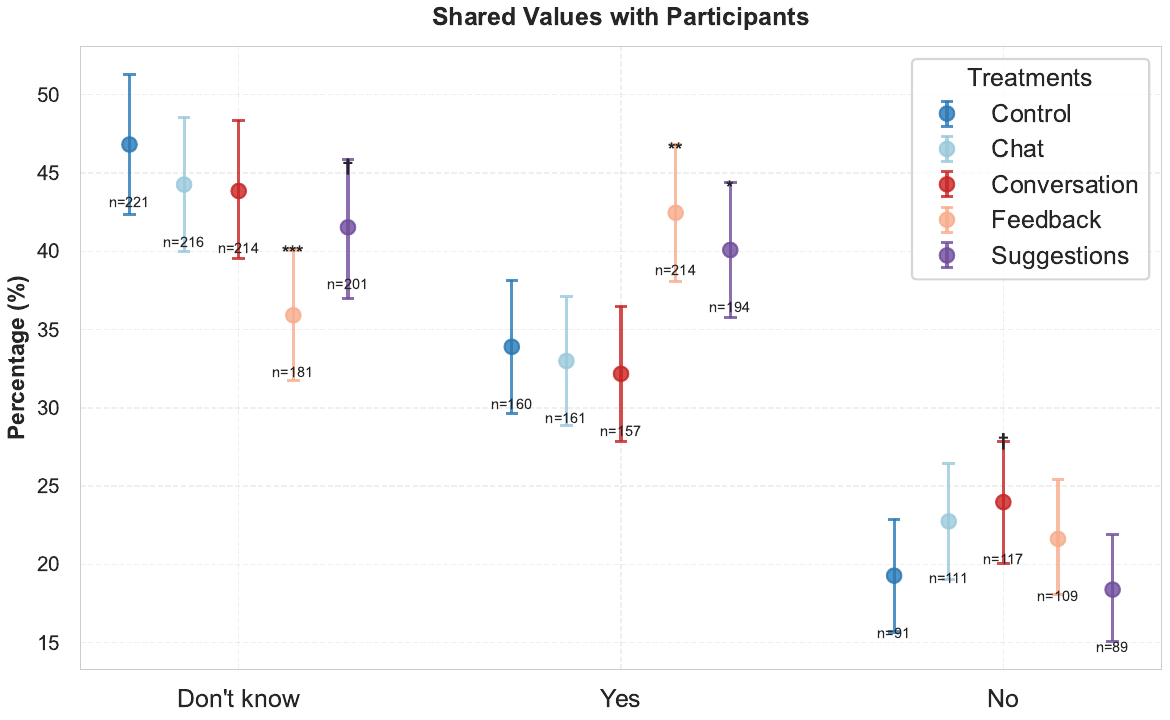}
    \end{minipage}
    
\end{figure}

\begin{figure}[H]
    \centering
    
    \begin{minipage}[t]{0.48\textwidth}
        \centering
        \includegraphics[width=\linewidth]{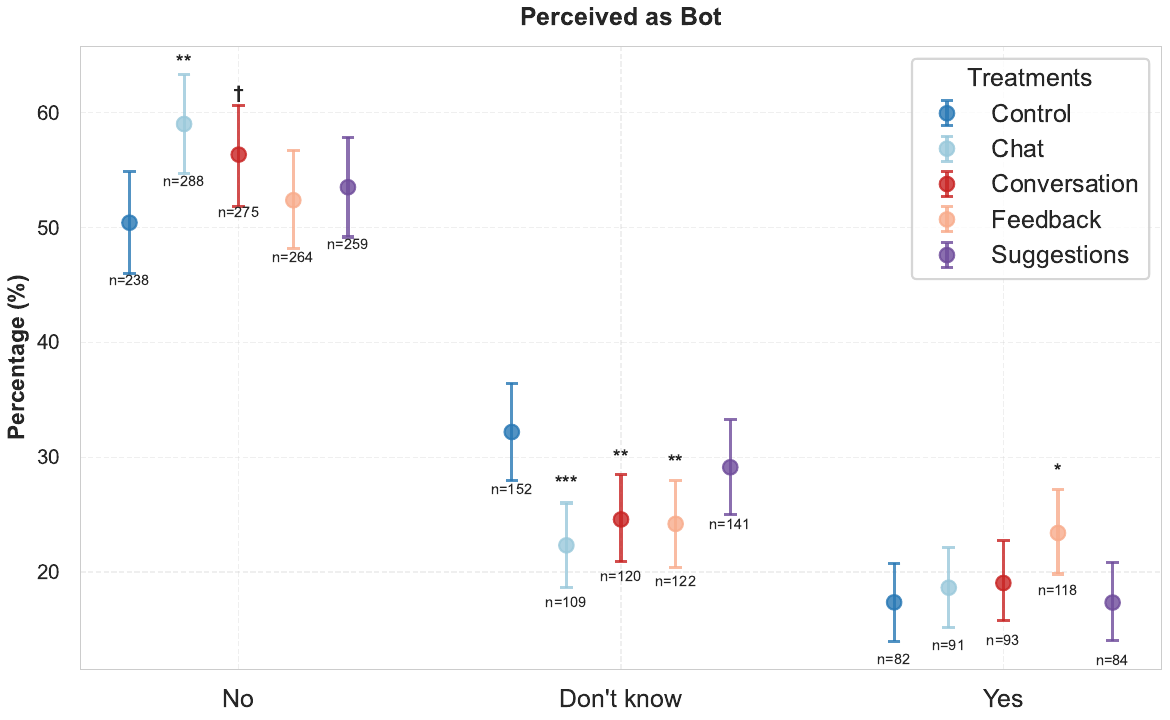}
    \end{minipage}%
    \hfill
    \begin{minipage}[t]{0.48\textwidth}
        \centering
        \includegraphics[width=\linewidth]{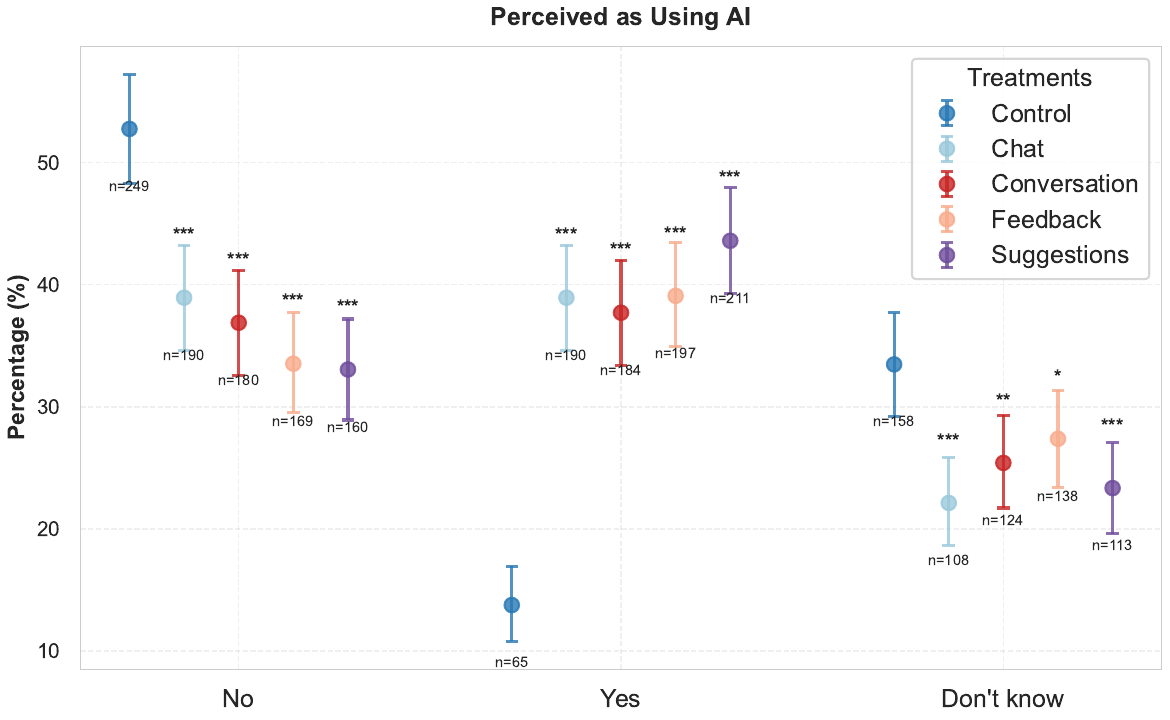}
    \end{minipage}%

    \label{fig:user-ratings}
    \caption{[\textit{Continued from last page}] Ratings of Users and Comments: Measuring user responses to comments received and other participants in study group.}
    
\end{figure}

\end{document}